\documentclass{bmcart}

\usepackage{amsthm,amsmath}
\usepackage[utf8]{inputenc} 
\usepackage{graphicx}
\usepackage{bm}
\usepackage{amsfonts}
\usepackage{threeparttable}
\usepackage{multirow}
\usepackage{subcaption}
\usepackage{booktabs}
\usepackage{amssymb}
\usepackage{url}
\usepackage{rotating}

\startlocaldefs
\endlocaldefs

\begin{document}

\begin{frontmatter}

\begin{fmbox}
\dochead{Research}


\title{A Multimodal Graph Neural Network Framework for Cancer Molecular Subtype Classification}


\author[
  addressref={aff1},                   
  email={bingjun.li@uconn.edu}   
]{\inits{B.L.}\fnm{Bingjun} \snm{Li}}
\author[
  addressref={aff1},
  corref={aff1},
  email={sheida.nabavi@uconn.edu}
]{\inits{S.N.}\fnm{Sheida} \snm{Nabavi}}


\address[id=aff1]{
  \orgdiv{Department of Computer Science and Engineering},             
  \orgname{University of Connecticut},          
  \city{Storrs},                              
  \cny{US}                                    
}



\end{fmbox}


\begin{abstractbox}

\begin{abstract} 
\parttitle{Background} 
The recent development of high-throughput sequencing creates a large collection of multi-omics data, which enables researchers to better investigate cancer molecular profiles and cancer taxonomy based on molecular subtypes. Integrating multi-omics data has been proven to be effective for building more precise classification models. Most current multi-omics integrative models use either an early fusion in the form of concatenation or late fusion with a separate feature extractor for each omic, which are mainly based on deep neural networks. Due to the nature of biological systems, graphs are a better structural representation of bio-medical data. Although few graph neural network (GNN) based multi-omics integrative methods have been proposed, they suffer from three common disadvantages. One is most of them use only one type of connection, either inter-omics or intra-omic connection; second, they only consider one kind of GNN layer, either graph convolution network (GCN) or graph attention network (GAT); and third, most of these methods have not been tested on a more complex classification task, such as cancer molecular subtypes.  

\parttitle{Results} 
In this study, we propose a novel end-to-end multi-omics GNN framework for accurate and robust cancer subtype classification. The proposed model utilizes multi-omics data in the form of heterogeneous multi-layer graphs, which combine both inter-omics and intra-omic connections from established biological knowledge. The proposed model incorporates learned graph features and global genome features for accurate classification. We test the proposed model on the Cancer Genome Atlas (TCGA) Pan-cancer dataset and TCGA breast invasive carcinoma (BRCA) dataset for molecular subtype and cancer subtype classification, respectively. The proposed model shows superior performance compared to four current state-of-the-art baseline models in terms of accuracy, F1 score, precision, and recall. The comparative analysis of GAT-based models and GCN-based models reveals that GAT-based models are preferred for smaller graphs with less information and GCN-based models are preferred for larger graphs with extra information.
\end{abstract}




\end{abstractbox}
%

\end{frontmatter}



\section*{Background}
The fast-growing high-throughput sequencing technology has made DNA and RNA sequencing more efficient and accessible, resulting in a large collection of multi-omics data which makes molecular profiling possible. Due to the heterogeneity in cancer and the complexity of the biological processes, employing multi-omics sequencing data are crucial to more accurate cancer classification and tumor profiling. Many researchers have proposed methods that incorporate multi-omics data for either cancer type classification or cell type clustering~\cite{Li2021,Zhang2019,Yang2019,sharifi2019,Wang2021,Ma2019,Kaczmarek2022,Lotfollahi2022,Huang2019,bai2022semi,chai2021integrating}. These methods show that utilizing multi-omics data improves performance, and provides a better understanding of the key pathophysiological pathways across different molecular layers~\cite{Heo2021}. A typical multi-omics data generated from DNA and RNA sequencing usually consists of mRNA expression, microRNA (miRNA) expression, copy number variation (CNV), and DNA methylation \cite{hoadley2014multiplatform}. The difference in data distributions across each omic, and the complex inter-omics and intra-omic connections (certain omic can act as a promotor or suppressor to genes) add more challenges to developing an integrative multi-omics classification method for cancer molecular subtypes.

Recent studies have shown that cancer taxonomy based on molecular subtypes can be crucial for precision oncology \cite{hoadley2014multiplatform, mateo2022delivering}.
An accurate cancer molecular subtype classifier is crucial for early-stage diagnosis, prognosis, and drug development. Traditional cancer taxonomy is based on its tissue origin. In 2014, The Cancer Genome Atlas (TCGA) Research Network proposed a new clustering method for cancers based on their integrated molecular subtypes that share mutations, copy-number alterations, pathway commonalities, and micro-environment characteristics instead of their tissue of origin~\cite{hoadley2014multiplatform}. They found 11 subtypes from 12 cancer types. In 2018, they applied the new taxonomy method to 33 cancer types and found 28 molecular subtypes~\cite{hoadley2018cell}. The new cancer taxonomy provides a better insight into the heterogeneous nature of cancer.

With the recent development in deep learning models, data-driven models benefit from the powerful feature extraction capability of deep learning networks in many fields~\cite{Defferrard2016,zou2019primer,he2020data,wang2021clustering,shi2022constraint,he2023robust,wang2021two}. Most multi-omics integrative models employ an early fusion approach that aggregates multi-omics data (mainly by concatenation)  and then applies a deep neural network as a feature extractor; or a late fusion approach that first extracts features from each omic by deep neural networks and then aggregates extracted features as inputs to the classification network. For efficient implementation of multi-omics integrative models, convolutional neural networks (CNNs) are widely used~\cite{nicora2020integrated}. 

Traditional deep neural networks are based on the assumption that the inner structure of the data is in Euclidean space~\cite{wu2020comprehensive}. Because of the complex interactions across many biological processes, such data structure is not a proper representation of bio-medical data, and researchers proposed graph-based data structures to tackle this limitation. In 2016, a graph convolution network (GCN), ChebNet, was proposed~\cite{Defferrard2016}. It uses the Chebyshev polynomial as the localized learning filter to extract the graph feature representation. In 2017, Petar Velickovic et al. proposed a graph attention network (GAT) that overcomes GCN's disadvantage of dependence on the Laplacian eigenbasis~\cite{petar2017}. GAT uses masked self-attention layers to enable nodes to attend over their neighborhoods' features~\cite{petar2017}. With the recent growing interest in the graph neural network, many graph-based classification methods have been proposed in the bio-medical field.

To utilize the power of graph-structured data, Ramirez et al. proposed a GCN method to use intra-omic connections, protein-protein interaction networks, and gene co-expression networks. The model achieves a 94.71\% classification accuracy for 33 cancer types and normal tissue on TCGA data \cite{ramirez2020classification}. To use the intra-omic connection across multiple omics, Wang et al. proposed MOGONET, a late-fusion GCN-based method that integrates multi-omics data for bio-medical data classification. And they achieve 80.61\% accuracy on breast cancer subtype classification with BRCA dataset~\cite{Wang2021}. To compensate for the limitation of GCN, that it only extracts local representation on the graph, Li et al. proposed a parallel-structured GCN-based method that utilizes a gene-based prior knowledge graph for cancer molecular subtype classification~\cite{Li2021}. There are also other ways to structure the graph. Wang et al. proposed a GCN-based method that uses a KNN-generated cell-cell similarity graph for single-cell sequencing data classification~\cite{wang2021single}. 

Since the introduction of GAT in 2017, it has gained more and more interest. Shanthamallu et al. proposed a GAT-based method, GrAMME, with two variations that use a supra-graph approach and late-fusion approach to extract features from a multi-layer graph with intra-omic connections only for classification in social science and political science datasets ~\cite{Shanthamallu2020}. On the other hand, Kaczmarek et al. proposed a multi-omics graph transformer to utilize an inter-omics connection only graph, the miRNA-gene target network, for cancer classification on 12 cancer types from the TCGA data~\cite{Kaczmarek2022}.

There are three common disadvantages of these approaches. First, most of them consider only one kind of connections in their model, either inter-omics or intra-omic connections. They do not aim to utilize both inter-omics and intra-omic connections for more effective feature extraction. Second, they only consider one kind of GNN models, either GCN or GAT. We find that GAT and GCN have their strength in different scenarios as shown in our experiments. Different graph layers are preferred for different tasks even with datasets in a similar domain. Third, most of these methods have not been tested on a more complex classification task. They are used for classification based on the cell-of-origin taxonomy such as cancer type classification and have not been applied to a more complex classification task such as cancer molecular subtype classification, which is more useful for diagnosis, prognosis, and treatment. Inspired by our previous work on the cancer molecular subtype classification based solely on intra-omic connections, we aim to develop a multi-omics integrative framework that exploits the powerful data aggregation property of  GCN or GAT models (depending on the situation) and utilizes both the intra-omic network and the inter-omics network for more precise classification.

Our goal is to build an accurate, robust, and efficient multi-omics integrative predictive model to classify these cancer molecular subtypes. 
In this work, we propose a general framework that can be used with any graph neural networks as the feature extractor, incorporate both gene-based and non-gene-based prior biological knowledge (primarily miRNA), and learn a knowledge graph consisting of both intra-omic and inter-omics connections. We apply the proposed model to classify cancer molecular subtypes and breast cancer molecular subtypes. We choose breast cancer as it is one of the most common and lethal cancers with a large number of samples in TCGA. It can be categorized into four major molecular subtypes based on the gene expression of the cancer cells, and breast cancer subtypes have significant impacts on the patient's survival rates ~\cite{onitilo2009breast}. Our experimental results show the proposed method outperforms both the graph-based and CNN-based state-of-the-art methods.

Our contributions in this study are i) a novel generalized GNN-based multi-omics integrative framework for cancer molecular subtype classification, ii) a supra-graph approach that can incorporate both intra-omic and inter-omics prior biological knowledge in the form of graphs, iii) a representation of multi-omics data in the form of heterogeneous multi-layer graph, and iv) a comparative analysis of GCN and GAT based models at different combinations of omics and different graph structures.

\begin{figure*}[htp]
\centering
   \includegraphics[width=0.78\textwidth]{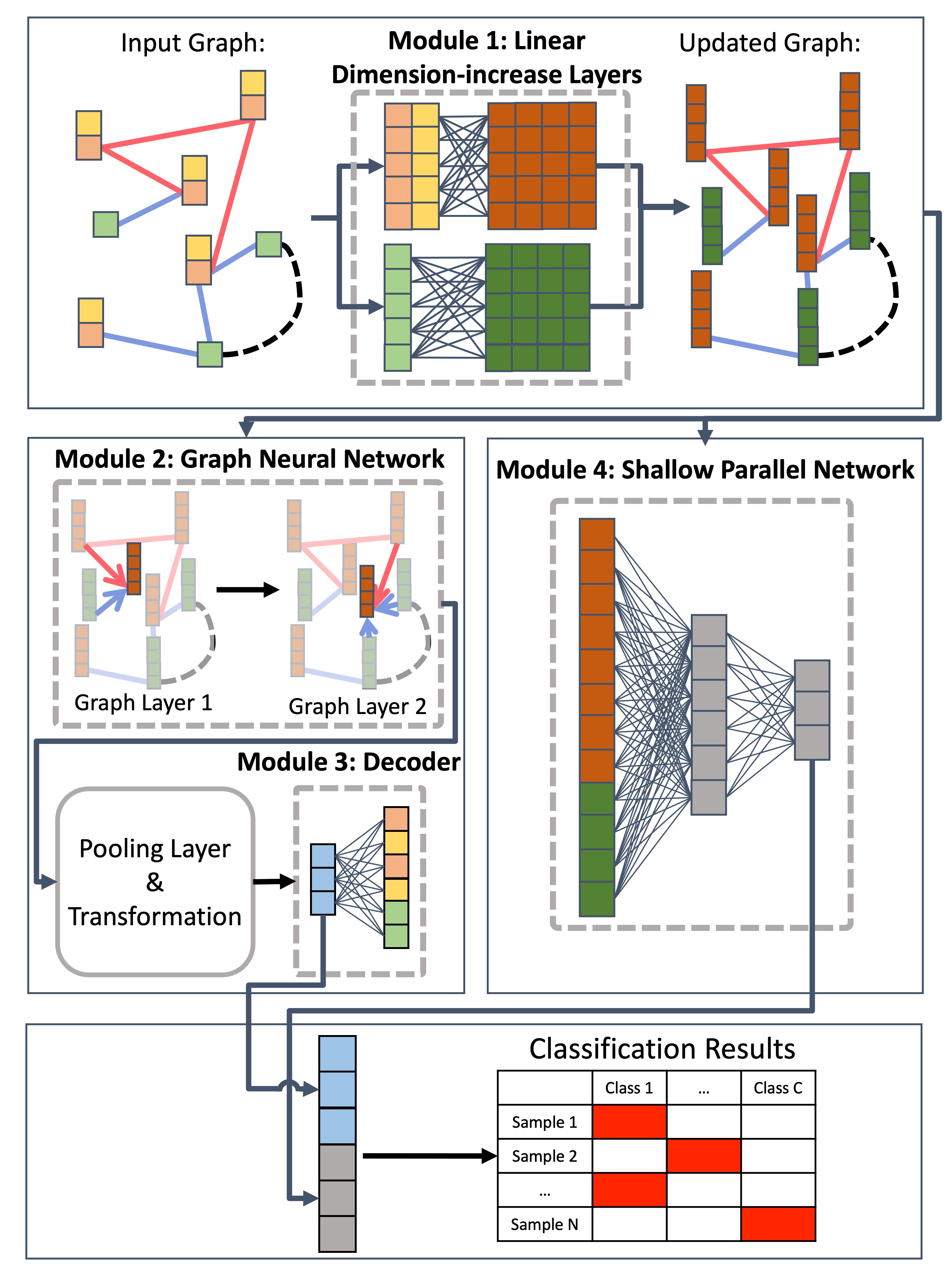}
    \caption{The overall structure of the proposed model has four major modules shown as dotted grey rectangles. The input graph consists of inter-omics (red edges), intra-omic (blue edges) edges and miRNA-miRNA meta-path (black dashed edges), and three omics data, mRNA (orange boxes), CNV (yellow boxes), and miRNA (green boxes) is shown as the leftmost side. Module 1 consists of two parallel linear dimension-increase layers for gene-based nodes and miRNA-based nodes. The upgraded graph shown in the middle is obtained by feeding the node attributes from the input graph through module 1, where the dark orange boxes are the updated gene-based node attributes and the dark green boxes  are the updated miRNA-based node attributes. Module 2 consists of two graph neural network layers, which can be any graph neural networks. The output of module 2 is then fed into a max pooling layer and then a transformation layer to obtain the learned graph representation (blue boxes). Module 3 consists of a decoder to reconstruct the graph representation back to the input graph node attributes. Module 4 consists of a shallow fully connected network that takes the updated node attributes as the input. The output of the parallel network (grey cubes) is then concatenated with the learned graph representation, and passes through a classification layer for the classification task.}
    \label{fig:model}
\end{figure*}

\section*{Method and Materials}

The overview of the proposed framework structure is shown in Figure \ref{fig:model}. The input data for the proposed framework is shown as a graph structure on the leftmost side. The data consists of three omics, mRNA expression (orange boxes), copy number variation (CNV) (yellow boxes), and miRNA (green boxes). The details of the network structure are discussed in the following Network Section. The proposed framework consists of 4 major modules: Module 1) a linear dimension-increase neural network, Module 2) a graph neural network (GNN), Module 3) a decoder, and Module 4) a shallow parallel network. Any kind of graph neural network can be used in Module 2. In this study, we focus on graph convolutional network (GCN) and graph attention network (GAT), which are two major kinds of GNN. 
Experiments about the effect of the decoder and the shallow parallel network  modules are discussed in our ablation study.

\begin{figure}[h]
\centering
   \includegraphics[width=0.9\textwidth]{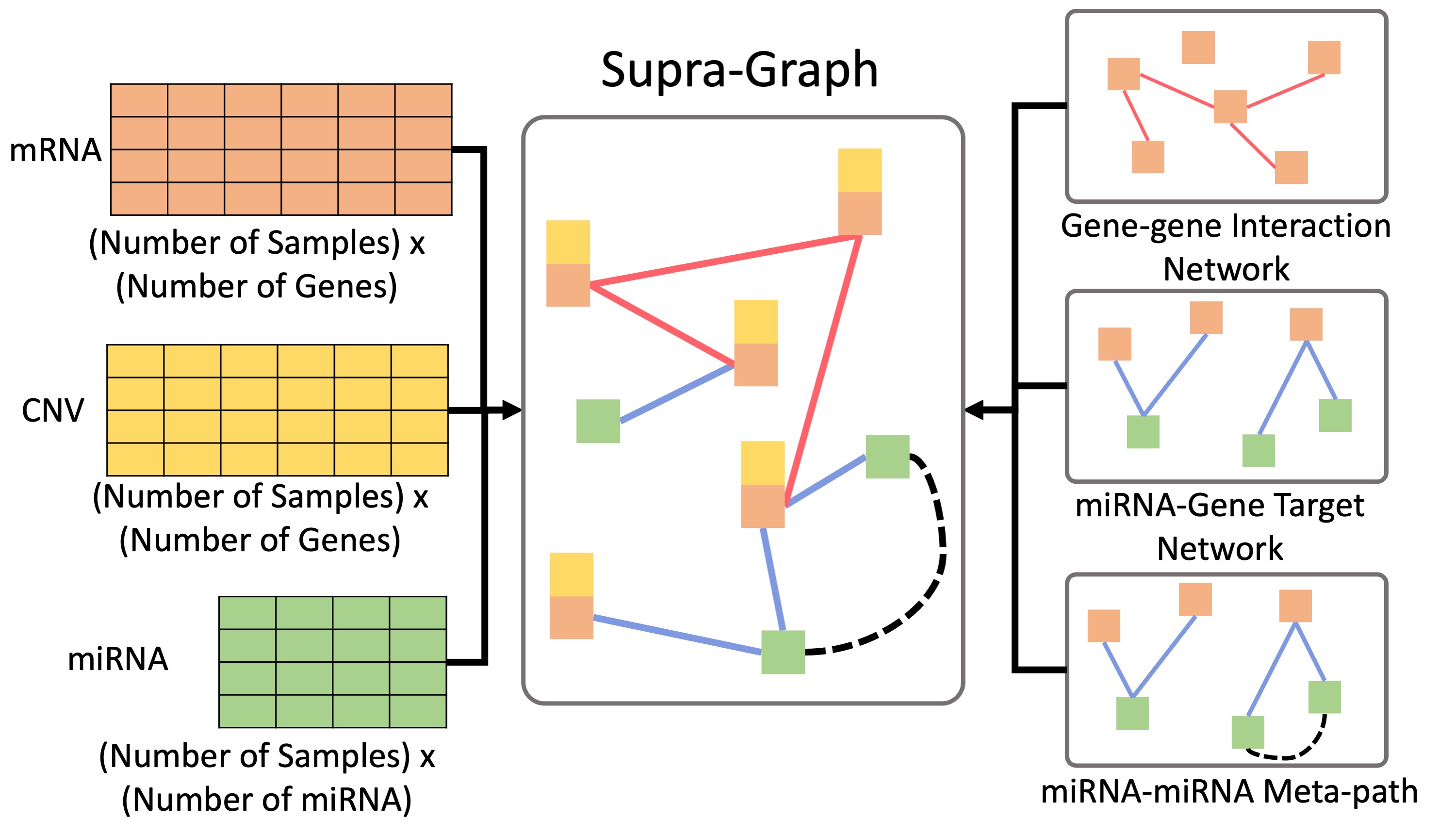}
    \caption{The overall graph, supra-graph, is constructed from three different omic data on the left-hand side and two prior knowledge graphs on the right-hand side. mRNA (orange table) and CNV (yellow table) data are considered gene-based, which have the same dimension. miRNA (green table) data has the same number of rows but different feature lengths for each sample.}
    \label{fig:network}
\end{figure}

\subsection*{Network}

We build a heterogeneous multi-layer graph based on the prior biological knowledge, i.e. gene-gene interaction (GGI) network from BioGrid and miRNA-gene target network from miRDB~\cite{pmid33070389,chen2020mirdb}. Inspired by the meta-path and supra-graph approach for the multi-layered network models~\cite{Lee2020,Shanthamallu2020}, we build a supra-graph with miRNA-miRNA meta-paths.  A miRNA-miRNA meta-path is defined as if two miRNA nodes are connected to the same gene node from the GGI network and miRNA-gene network. An example of how we construct the supra-graph is shown in Figure \ref{fig:network}. Meta-paths are shown as dotted lines in the figure.

The adjacency matrix of the supra-graph is an $(N + M) \times (N + M)$ matrix, where $N$ is the number of genes and $M$ is the number of miRNA. Every node in the graph is assumed to be self-connected, thus the diagonal elements of  the adjacency matrix in the study are 1. The adjacency matrix of the supra-graph is shown in Equation \eqref{eq:adj}.

\begin{equation}
\textbf{A}_{Supra} = \begin{bmatrix}
\textbf{A}_{gene-gene} & \textbf{A}_{gene-mi}\\
\textbf{A}_{gene-mi}^T & \textbf{A}_{mi-mi},
\end{bmatrix}\label{eq:adj},
\end{equation}
 where $\textbf{A}_{gene-gene} \in \mathbb{R}^{N \times N}$, $\textbf{A}_{gene-mi} \in \mathbb{R}^{N \times M}$, and $\textbf{A}_{mi-mi} \in\mathbb{R}^{M \times M}$.

We also construct four different kinds of graphs other than supra-graph in our ablation study and apply them to  five input combinations of omics: mRNA, miRNA, mRNA + miRNA, mRNA+CNV, mRNA + MiRNA + CNV,  to test the effect of the different graphs on the model performance. The four different graphs are defined as follows.

\textbf{Only Gene-based Nodes}: When the input combination of omics is mRNA or mRNA+mRNA+CNV ($M=0$), the  graph is built with the GGI network, $\textbf{A} = \textbf{A}_{gene-gene} \in \mathbb{R}^{N \times N}$.

\textbf{Only miRNA-based Nodes}: When the input combination of omics is miRNA ($N=0$), the  graph  is built with only miRNA meta-path network, $\textbf{A} = \textbf{A}_{mi-mi} \in \mathbb{R}^{M \times M}$.

\textbf{Only Intra-class Edges}: The  graph only contains GGI network and miRNA meta-path network. 
\begin{equation}
    \textbf{A}_{Supra} = \begin{bmatrix}
    \textbf{A}_{gene-gene} & \bm{0}_{N,M}\\
    \bm{0}_{M,N} & \textbf{A}_{mi-mi}
    \end{bmatrix} \in \mathbb{R}^{(N+M) \times (N+M)}.
\end{equation}

\textbf{Only Inter-class Edges}: The  graph only contains miRNA-gene target network. 
\begin{equation}
    \textbf{A}_{Supra} = \begin{bmatrix}
    \textbf{I}_{N,N} & \textbf{A}_{gene-mi}\\
    \textbf{A}_{gene-mi}^T & \textbf{I}_{M,M}
    \end{bmatrix} \in \mathbb{R}^{(N+M) \times (N+M)}.
\end{equation}

The input graph is denoted as a tuple $\mathcal{G}=(V,E,\textbf{X}_V)$, where $V$ is the set of nodes, $E$ is the set of edges, and $\bm{x}_V$ is the node attributes. The prior knowledge is incorporated into the model through the supra-graph defined  above. In the supra-graph, nodes consist of both gene-based nodes and miRNA-based nodes, and edges are assigned by the adjacency matrix. Each gene-based node has a node attribute of a vector consisting of both gene expression and CNV data, $\bm{x}_{v \in V_{gene}} \in \mathbb{R}^{2}$. Each miRNA-based node has a node attribute as a scalar, $x_{v \in V_{\text{miRNA}}} \in \mathbb{R}$. The gene-based nodes and miRNA-based nodes are fed through a linear dimension-increase layer, denoted as Module 1 in Figure \ref{fig:model} to achieve the same node attribute dimension, $\textbf{X}'_V \in \mathbb{R}^{(N+M) \times F}$, where $F$ is the increased node attribute dimension.

\subsection*{Graph Neural Network: Convolution-based}
As mentioned before, any graph neural network can be used in the GNN module. We use ChebNet  \cite{Defferrard2016} to implement the GCN in this study. The supra-graph adjacency matrix introduced  in the previous network section is first Laplacian normalized to $\textbf{L}$ as expressed in Equation \eqref{eq:Laplacian}.
\begin{equation}
\label{eq:Laplacian}
    \textbf{L} = \textbf{I}+\textbf{D}^{-1/2}\textbf{A}\textbf{D}^{1/2},
\end{equation}
where $\textbf{I} \in \mathbb{R}^{(N+M) \times (N+M)}$ is an identity matrix, and the degree matrix $\textbf{D} \in \mathbb{R}^{(N+M) \times (N+M)}$ is a diagonal matrix. The eigen decomposition form of $\textbf{L}$ can be obtained as 
\begin{equation}
\label{eq:laplacian_decomposition}
\textbf{L}=\textbf{U} \Lambda \textbf{U}^T,
\end{equation}
where $\textbf{U}=(\bm{u}_1, \bm{u}_2, \dots, \bm{u}_n)$ is a matrix of $n$ orthonormal eigenvectors of $\textbf{L}$, therefore $\textbf{UU}^T=\textbf{I}$. And $\boldsymbol{\Lambda} = diag(\lambda_1, \lambda_2, \dots, \lambda_n)$ is the eigenvalue matrix \cite{Defferrard2016}.

After transforming the graph on the Fourier domain, the learning filter can be approximated by a $K^{th}$-order Chebshev polynomial. The convolution on the graph by such localized learning filter, $h(\boldsymbol{\Lambda})$ can be expressed in Equation \eqref{eq:gcn_learning_filter}.
\begin{equation}\label{eq:gcn_learning_filter}
    y =  \textbf{U} h(\boldsymbol{\Lambda}) \textbf{U}^T \textbf{X}_j = \textbf{U}\sum_{k=1}^{K-1} \beta_k T_k(\tilde{\boldsymbol{\Lambda}}) \textbf{U}^T \textbf{X}_j= \sum_{k=0}^{K-1}\beta_k T_k(\tilde{\textbf{L}}\textbf{X}_j),
\end{equation}
where $\textbf{X}_j \in \mathbb{R}^{(N+M) \times F}$ is the features of $j$-th sample, $\tilde{\textbf{L}}=2\textbf{L}/\lambda_{max}-\textbf{I}$, and $T_k(\tilde{\textbf{L}})=2\tilde{\textbf{L}}T_{k-1}(\tilde{\textbf{L}})-T_{k-2}(\tilde{\textbf{L}})$ with $T_0(\tilde{\textbf{L}})=\textbf{I}$ and $T_1(\tilde{\textbf{L}})=\tilde{\textbf{L}}$. $K$ is a hyper-parameter, where $K=5$ in our study. A max-pooling layer with $p=8$ is used to reduce the number of nodes and one layer of fully connected network is used to transform the learned local feature representation to a vector of length 64 for each sample, $\boldsymbol{\theta}_1 \in \mathbb{R}^{64}$.

\subsection*{Graph-Neural Network: Attention-based}

GAT aims to solve the problem of GCN's dependence on Laplacian eigenbasis of the graph adjacency matrix~\cite{petar2017}. The updated node attributes are first passed through a linear transformation by a learnable weight, denoted as $\textbf{W} \in \mathbb{R}^{F' \times F}$, where $F$ is the updated node attribute dimension and $F'$ is the intended output dimension for this GAT layer. Then, the self-attention coefficients for each node can be calculated as Equation \eqref{eq:attention-coefficent}.
\begin{equation}\label{eq:attention-coefficent}
    e_{ij}=a(\textbf{W} \bm{x}_i, \textbf{W} \bm{x}_j),
\end{equation}
where $e_{ij}$ represents the importance of node $j$ to node $i$ and $\bm{x}_i, \bm{x}_j$ are the node attributes for node $i,j$. Such attention score is only calculated for $j \in NB(i)$, where $NB(i)$ is all the first-order neighbor nodes around node $i$. The method normalizes the attention score by a softmax layer of $e_{ij}$ and uses LeakyReLU as the activation function as express in Equation \eqref{eq:normalized-attention-coefficents}. 
\begin{equation}\label{eq:normalized-attention-coefficents}
    \alpha_{ij} = \frac{\exp(\text{LeakyReLU}(\vec{\bm{a}}^{\,T}[\textbf{W}\bm{x}_i||\textbf{W}\bm{x}_j]))}{\sum_{k \in NB(i)}\exp(\text{LeakyReLU}(\vec{\bm{a}}^{\,T}[\textbf{W}\bm{x}_i||\textbf{W}\bm{x}_k]))}
\end{equation}
The output for each node can be expressed as  Equation \eqref{eq:gat_output}.
\begin{equation}\label{eq:gat_output}
    \bm{x}'_i = \sigma (\sum_{j \in NB(i)}\alpha_{ij} \textbf{W} \bm{x}_j).
\end{equation}
A multi-head attention mechanism is used to stabilize the attention score. In our study, the number of heads is 8. Similar to the GCN-based GNN module, the output is then passed through a max-pooling layer and a transformation layer to obtain the local graph representation, $\boldsymbol{\theta}_1 \in  \mathbb{R}^{64}$.

\subsection*{Decoder \& Shallow Parallel Network}

As shown in Figure \ref{fig:model}, the decoder is a two-layer fully connected network that is used to reconstruct the node attributes on the input graph. To compensate the localization property of either GCN or GAT layer in the GNN module, we use a parallel shallow fully connected network. Since the prior knowledge graphs have many limitation~\cite{Li2021},  we may  neglect some global patterns in the data when extracting features based on the graph structure only. A shallow two-layer fully connected network is able to learn the global features of the data while ignoring the actual inner structure of the data. These two modules help the framework to better extract the overall sample feature representation. The effect of including vs. excluding these two modules is discussed in detail in the Ablation Study Section.

The input of the parallel network is the updated node attributes, $\textbf{X}'_V \in \mathbb{R}^{(N+M) \times F}$ and the output global representation of the sample, $\boldsymbol{\theta}_1$ is in the same dimension as the local feature representation from the GNN module, $\boldsymbol{\theta}_2 \in \mathbb{R}^{64}$. $\boldsymbol{\theta}_1$ and $\boldsymbol{\theta}_2$ are then concatenated and passed through a classification layer for prediction.

\subsection*{Loss Function}
\label{subsection:lossfunction}
In the proposed framework, we define the loss function $L$ as a linear combination of three loss functions in Equation \eqref{eq:loss}.
\begin{equation}
\label{eq:loss}
    L = \lambda_1 L_{ent} + \lambda_2 L_{recon} + \lambda_3 L_{reg},
\end{equation}
where $\lambda_1$, $\lambda_2$ and $\lambda_3$ are linear weights, $L_{ent}$ is the standard cross-entropy loss for the classification results, $L_{recon}$ is the mean squared error for the reconstruction loss when the decoder is included, and $L_{reg}$ is the squared $l^2$ norm of the model parameters to penalize the number of parameters to avoid overfitting. $L_{recon}$ is defined as 
\begin{equation}L_{recon} =\sum_{j}(\bm{x}_{j}-\hat{\bm{x}}_{j})^2,\end{equation}
where $\bm{x}_{j}$ is the flattened feature vector of $j$-th sample and $\hat{\bm{x}}_{j}$ is the corresponding reconstructed vector. We denote $\textbf{W}_{all}$ as the vector consists of all parameters in the model and the $L_{reg}$ is defined as \begin{equation}L_{reg} = \sum_{w \in \textbf{W}_{all}} w^2.\end{equation}

\section*{Results and Discussion}

We apply the proposed model to two different classification problems. The first is  cancer molecular subtype classification on the TCGA Pan-cancer dataset and the second is  breast cancer subtype classification on the TCGA breast invasive carcinoma (BRCA) dataset~\cite{hoadley2018cell, cancer2012comprehensive}.

\subsection*{Data and Experiment Settings}

\begin{figure}[h]
\centering
   \includegraphics[width=0.6\textwidth]{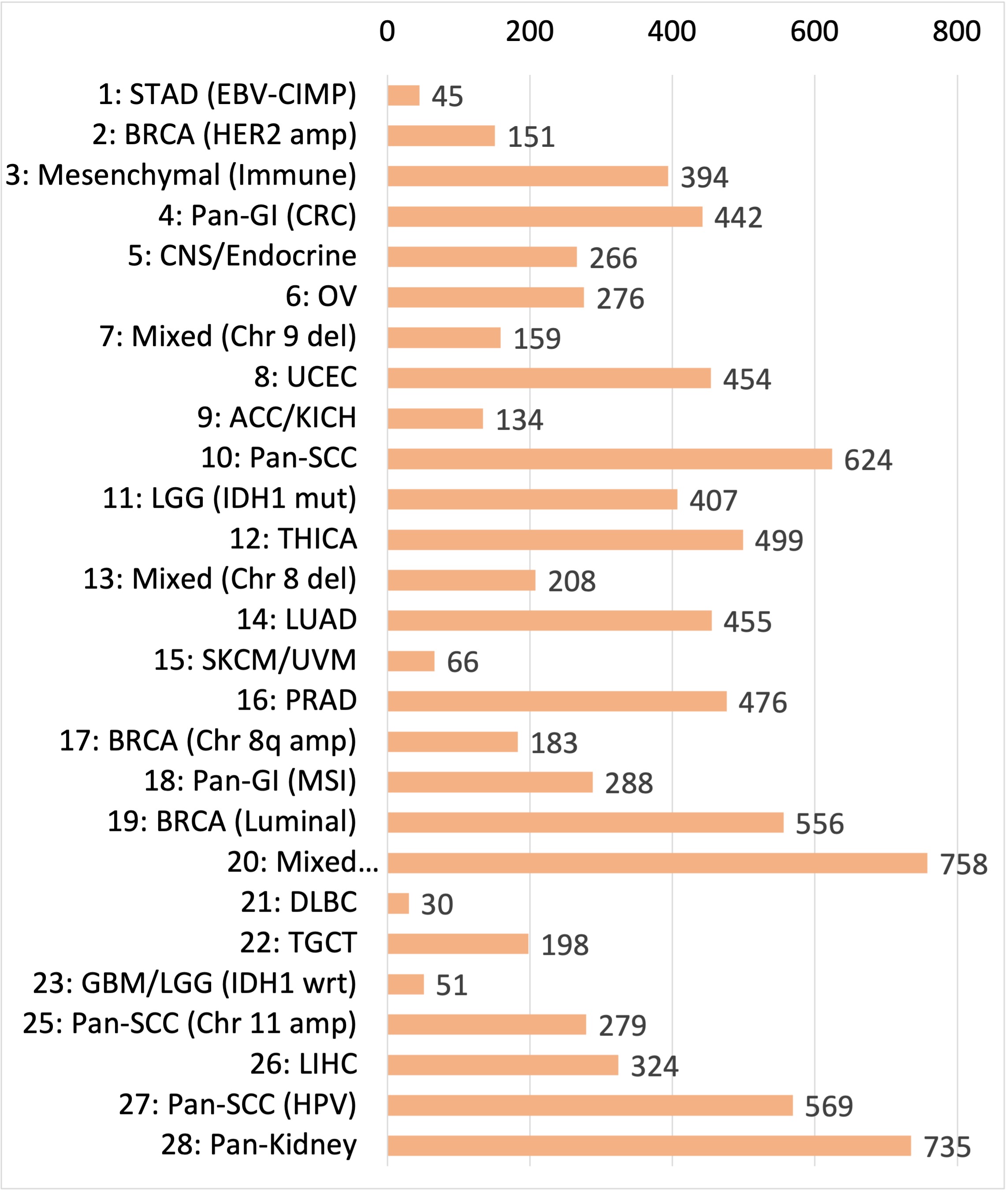}
    \caption{The number of cases in each molecular subtypes is shown. All samples from class 24 are excluded due to lack of miRNA data.}
    \label{fig:tcga}
\end{figure}

The TCGA Pan-cancer RNA-seq data, CNV data, miRNA data, and molecular subtype labels are obtained from the University of California Santa Cruz's Xena website~\cite{goldman2020visualizing}. We only keep samples that have all three omics data and molecular subtype labels, and collect 9,027 samples in total. We use 17,946 genes that are common in both the gene expression data and the CNV data, and 743 miRNAs. The total number of molecular subtypes is 27 and there is a clear imbalance among these 27 classes as shown in Figure \ref{fig:tcga}. All samples from class 24 are excluded from the study due to the lack of miRNA data. For BRCA subtype classification, there are 981 samples in total with 4 subtypes as shown in Table \ref{tab:brca}. For the experiments on both datasets, 80\% of the data is used for training, 10\% is used for validation, and 10\% is used for testing. All classes are present in the test set.

\begin{table}[htp]
\centering
\caption{Number of Cases in Each BRCA Subtype}
\label{tab:brca}
    \resizebox{0.3\textwidth}{!}{
    \begin{threeparttable}
    \begin{tabular}{ll}
    \hline
    BRCA Subtypes & Counts \\
    \hline
    LumA          & 529    \\
    LumB          & 197    \\
    Basal         & 175    \\
    Her2          & 80    \\
    \hline
    \end{tabular}
\end{threeparttable}
}
\end{table}

All expression values are normalized within their own omics. We select the top 700 genes ranked by gene expression variances across the samples, and the top 100 miRNAs by miRNA expression variance. Results are averaged from five individual trials. The details of the model structure and hyperparameters are disclosed in the appendix. The model is implemented using Pytorch Geometric Library.

\subsection*{Baseline Models}
We selected four state-of-the-art models \cite{Li2021, Kaczmarek2022, ramirez2020classification, Shanthamallu2020} as baseline models to evaluate the performance of the proposed approach. These four baseline models are implemented within the proposed framework in two forms, one is with the original structure, and the other is with some modifications to accommodate the multi-omics data. The details of all graph-based baseline implementation configurations are shown in Table \ref{tab:baseline}. We also included a fully-connected neural network (FC-NN) as a Euclidean-based baseline model. Conventional machine learning methods, such as Random Forest and SVM are not included in the scope of this study because they do not scale well to the multi-omics data as mentioned in our previous work~\cite{Li2021}. 

\begin{table*}[htp]
\centering
\caption{Configurations of Baseline Models on Omics, Graph Structure, GNN Layers, and Regularizaiton Modules}
\label{tab:baseline}
\resizebox{\textwidth}{!}{
\begin{threeparttable}
\begin{tabular}{l|ccc|cc|cc|cc}
\hline
\multirow{2}{*}{Model}       & \multicolumn{3}{c|}{Omics}                                                         & \multicolumn{2}{c|}{Graph}                             & \multicolumn{2}{c|}{GNN Layer}                         & \multicolumn{2}{c}{Module}                            \\
\cline{2-10}
                           & mRNA                      & CNV                       & miRNA                     & Intra-omic                & Inter-omic                & GCN                       & GAT                       & Decoder                   & Parallel                  \\
\hline
GCN (Original) \cite{ramirez2020classification}             & \checkmark &              --             &               --            & \checkmark &               --            & \checkmark &            --               &             --              &               --            \\
GCN (Modified)             & \checkmark & \checkmark & \checkmark & \checkmark &         --                  & \checkmark &        --                   &               --            &             --              \\
Multi-omics GCN (Original) \cite{Li2021} & \checkmark & \checkmark &          --                 & \checkmark &          --                 & \checkmark &            --               & \checkmark & \checkmark \\
Multi-omics GCN (Modified) & \checkmark & \checkmark & \checkmark & \checkmark &          --                 & \checkmark &          --                 & \checkmark & \checkmark \\
GrAMME (Modified) \cite{Shanthamallu2020}        & \checkmark & \checkmark & \checkmark &      --                     & \checkmark &          --                 & \checkmark &            --               &         --                  \\
Multi-omics GAT (Original) \cite{Kaczmarek2022} & \checkmark &           --                & \checkmark & \checkmark &            --               &             --              & \checkmark &            --               &            --               \\
Multi-omics GAT (Modified) & \checkmark & \checkmark & \checkmark & \checkmark &           --                &              --             & \checkmark &             --              &              --            \\
\hline
\end{tabular}
\end{threeparttable}
}
\end{table*}

\subsubsection*{Fully-connected Neural Network (FC-NN)}
The FC-NN is one of the widely used deep learning model for data in Euclidean space. The implemented structure is the same as the parallel structure. The input data is passed through a dimension-increase layer and then flattened. The flattened data is passed through three hidden layers and a softmax layer for classification. 

\subsubsection*{GCN Models by Ramirez et. al.}
The GCN model on cancer type classification is designed for gene expression data with intra-omic connections only~\cite{ramirez2020classification}. The implementation of the original structure and the modified structure is a GCN model with no regularization modules.

\subsubsection*{Multi-omics GCN Models by Li et al.}
The multi-omics GCN model on cancer molecular subtype classification is designed for gene expression and CNV data with intra-omic connections only~\cite{Li2021}. The implementation of both structures is a GCN model with a decoder and a parallel structure as shown in Table \ref{tab:baseline}.

\subsubsection*{GrAMME}
Since GrAMME is not designed for cancer type classification~\cite{Shanthamallu2020}, we modified the original structure for multi-omics data. GrAMME is designed for a GAT model with intra-omic connections only. The implementation is a GAT model with no regularization modules.

\subsubsection*{Multi-omics GAT by Kaczmarek et al.}
The multi-omics graph transformer on 12 cancer type classification is designed for gene expression and miRNA data with inter-omics connections only~\cite{Kaczmarek2022}. As shown in Table \ref{tab:baseline}, the main difference between multi-omics GAT and GrAMME is the construction of the graph.

\subsection*{Performance on Classification}

\begin{table*}[htp]
\centering
\caption{Results of the Proposed and Baseline Models with 700 Genes for Molecular Subtype Classification on the TCGA Pan-cancer Dataset And Cancer Subtype Classificaiton on the TCGA BRCA Dataset}
\label{tab:full_model_performance}
\resizebox{\textwidth}{!}{
\begin{threeparttable}
\begin{tabular}{l|ll|ll}
\hline
\multirow{2}{*}{Model}     & \multicolumn{2}{c|}{Pan-cancer}                                                         & \multicolumn{2}{c}{BRCA} \\
                           & Accu.\tnote{1} & F1 & Accu.\tnote{1}      & F1  \\ \hline
Proposed w/ GAT            & \textbf{83.9\%} $\bm{\pm}$ \textbf{1.4\%} & \textbf{0.84} $\bm{\pm}$ \textbf{0.01} & \textbf{86.4\%} $\bm{\pm}$ \textbf{1.9\%} & \textbf{0.87} $\bm{\pm}$ \textbf{0.02} \\
Proposed w/ GCN            & 81.2\% $\pm$ 0.6\% & 0.81 $\pm$ 0.01 & 83.8\% $\pm$ 0.9\% & 0.84 $\pm$ 0.01 \\
FC-NN                      & 78.4\% $\pm$ 0.8\% & 0.75 $\pm$ 0.02 & 80.8\% $\pm$ 1.1\% & 0.80 $\pm$ 0.02 \\
GCN (Original) \cite{ramirez2020classification}             & 77.6\% $\pm$ 0.9\% & 0.76 $\pm$ 0.02 & 82.8\% $\pm$ 1.2\% & 0.84 $\pm$ 0.01 \\
GCN (Modified)             & 78.5\% $\pm$ 1.2\% & 0.77 $\pm$ 0.02 & 81.8\% $\pm$ 1.4\% & 0.82 $\pm$ 0.01 \\
Multi-omics GCN (Original) \cite{Li2021} & 78.6\% $\pm$ 0.9\% & 0.78 $\pm$ 0.01 & 81.8\% $\pm$ 1.1\% & 0.82 $\pm$ 0.01 \\
Multi-omics GCN (Modified) & 80.2\% $\pm$ 0.8\% & 0.79 $\pm$ 0.01 & 82.8\% $\pm$ 0.9\% & 0.83 $\pm$ 0.01 \\
GrAMME (Modified) \cite{Shanthamallu2020}         & 81.4\% $\pm$ 1.3\% & 0.81 $\pm$ 0.03 & 82.8\% $\pm$ 1.6\% & 0.84 $\pm$ 0.03 \\
Multi-omics GAT (Original) \cite{Kaczmarek2022} & 76.3\% $\pm$ 1.2\% & 0.76 $\pm$ 0.02 & 81.8\% $\pm$ 1.3\% & 0.82 $\pm$ 0.02 \\
Multi-omics GAT (Modified) & 79.7\% $\pm$ 1.3\% & 0.79 $\pm$ 0.02 & 82.8\% $\pm$ 1.4\% & 0.84 $\pm$ 0.02 \\ \hline
\multicolumn{5}{l}{\footnotesize\textit{The bold font indicates the highest values and the values after $\pm$ sign are the standard deviations.}}\\
\end{tabular}
\begin{tablenotes}
\item[1] Accu. stands for Accuracy.
\end{tablenotes}
\end{threeparttable}
}
\end{table*}

For both classification tasks, the results of the proposed model and the baseline models are shown in Table \ref{tab:full_model_performance}. The proposed model with GAT layers outperforms all the baseline models for both tasks in all four metrics and the proposed model with GCN layers achieves third for the pan-cancer classification, and second for the breast cancer subtype classification. For the task of pan-cancer molecular subtype classification, the additional omic data in the modified structure improve the model performance in all three cases of the baseline model with the original structure vs. the baseline model with the modified structure. For the same task, the multi-omics GCN model with the decoder and parallel structure shows superior performance among all the baseline models that utilize GCN layers. And GrAMME, which utilizes intra-omic connections, performs better than GAT models that utilize inter-omics connections. GrAMME is the best-performing one among the baseline models for the pan-cancer task. Overall, we see the proposed model achieves the best performance for the classification task on the complex pan-cancer molecular subtype classification in all four metrics and we can conclude that more omics improve the performance of models, and the models with more restriction modules or GAT layers have better performance.

For breast cancer subtype classification, the overall trend is slightly different from that in the previous task. In most cases of including more omics, the performance of the models shows little or no improvement. We believe it is due to the nature of breast cancer taxonomy. The subtype is based on the expression level of multiple proteins. Thus, it makes the breast cancer subtype to be more closely related to the gene expression omic than the pan-cancer molecular subtype does. Such characteristic of the breast cancer subtype makes the model only using gene expression data perform very well such as the original GCN model. However, the proposed model still outperforms any baseline models by a large margin in all four metrics.

\subsection*{Ablation Study}
We conduct an ablation study to evaluate the effects of different numbers of genes, different training set splits, different combinations of modules within the model, and different combination of omics and graphs on the performance of the proposed model.

\subsubsection*{Different Numbers of Genes}

\begin{table*}[htp]
\centering
\caption{Results of the Proposed Model and Baseline Models with 300 and 500 Genes for Molecular Subtype Classification Using the TCGA Pan-cancer Dataset}
\label{tab:pan_cancer_ablation}
\resizebox{\textwidth}{!}{
\begin{threeparttable}
\begin{tabular}{l|ll|ll}
\hline
\multirow{2}{*}{Model}    & \multicolumn{2}{c|}{300}                                                               & \multicolumn{2}{c}{500}                                                                \\
                           & Accu.\tnote{1} & F1 & Accu. & F1 \\ \hline
Proposed w/ GAT            & \textbf{77.6\%} $\bm{\pm}$ \textbf{1.6\%} & \textbf{0.76} $\bm{\pm}$ \textbf{0.02} & \textbf{81.6\%} $\bm{\pm}$ \textbf{1.2\%} & \textbf{0.80} $\bm{\pm}$ \textbf{0.01} \\
Proposed w/ GCN            & 75.8\% $\pm$ 1.1\% & 0.74 $\pm$ 0.02 & 80.0\% $\pm$ 1.2\% & 0.79 $\pm$ 0.02 \\
FC-NN                      & 65.9\% $\pm$ 1.3\% & 0.59 $\pm$ 0.04 & 77.5\% $\pm$ 1.4\% & 0.74 $\pm$ 0.02 \\
GCN (Original)             & 74.5\% $\pm$ 1.6\%  & 0.72 $\pm$ 0.05 & 76.1\% $\pm$ 1.3\% & 0.73 $\pm$ 0.03 \\
GCN (Modified)             & 75.5\% $\pm$ 1.4\% & 0.72 $\pm$ 0.03 & 77.9\% $\pm$ 1.1\% & 0.77 $\pm$ 0.02 \\
Multi-omics GCN (Original) & 76.4\% $\pm$ 1.3\% & \textbf{0.76} $\bm{\pm}$ \textbf{0.03} & 77.4\% $\pm$ 1.3\% & 0.77 $\pm$ 0.03 \\
Multi-omics GCN (Modified) & 77.4\% $\pm$ 1.3\% & \textbf{0.76} $\bm{\pm}$ \textbf{0.02} & 78.2\% $\pm$ 1.2\% & 0.75 $\pm$ 0.02 \\
GrAMME (Modified)          & 77.4\% $\pm$ 1.5\% & \textbf{0.76} $\bm{\pm}$ \textbf{0.02} & 79.6\% $\pm$ 1.4\% & 0.79 $\pm$ 0.02 \\
Multi-omics GAT (Original) & 73.4\% $\pm$ 1.8\% & 0.71 $\pm$ 0.04 & 75.1\% $\pm$ 1.5\% & 0.74 $\pm$ 0.04\\
Multi-omics GAT (Modified) & 75.8\% $\pm$ 1.5\% & 0.74 $\pm$ 0.04 & 77.4\% $\pm$ 1.3\% & 0.74 $\pm$ 0.02 \\ \hline
\multicolumn{5}{l}{\footnotesize\textit{The bold font indicates the highest values and the values after $\pm$ sign are the standard deviations.}}\\
\end{tabular}
\begin{tablenotes}
\item[1] Accu. stands for Accuracy.
\item[2] Prec. stands for Precision.
\end{tablenotes}
\end{threeparttable}
}
\end{table*}

We trained the proposed model and all baseline models at the 300 and 500 genes for pan-cancer molecular subtype classification and 300, 500, 1000, 2000, and 5000 genes for breast cancer subtype classification. The limitation of the test scope on pan-cancer classification is due to the computation constraints caused by its large number of samples. As shown in Table \ref{tab:pan_cancer_ablation}, increasing the number of gene nodes improves the performance of all models. FC-NN model demonstrates great improvement in performance as the number of genes increases. And the proposed model with the GAT layer outperforms the baseline models at both numbers of genes.

\begin{figure*}[htp]
\caption{Performance of the Proposed Models and Baseline Models with Different Numbers of Genes on BRCA Dataset}
\label{fig_brca_result}
\begin{subfigure}{0.9\textwidth}
\centering
   \includegraphics[width=0.95\textwidth]{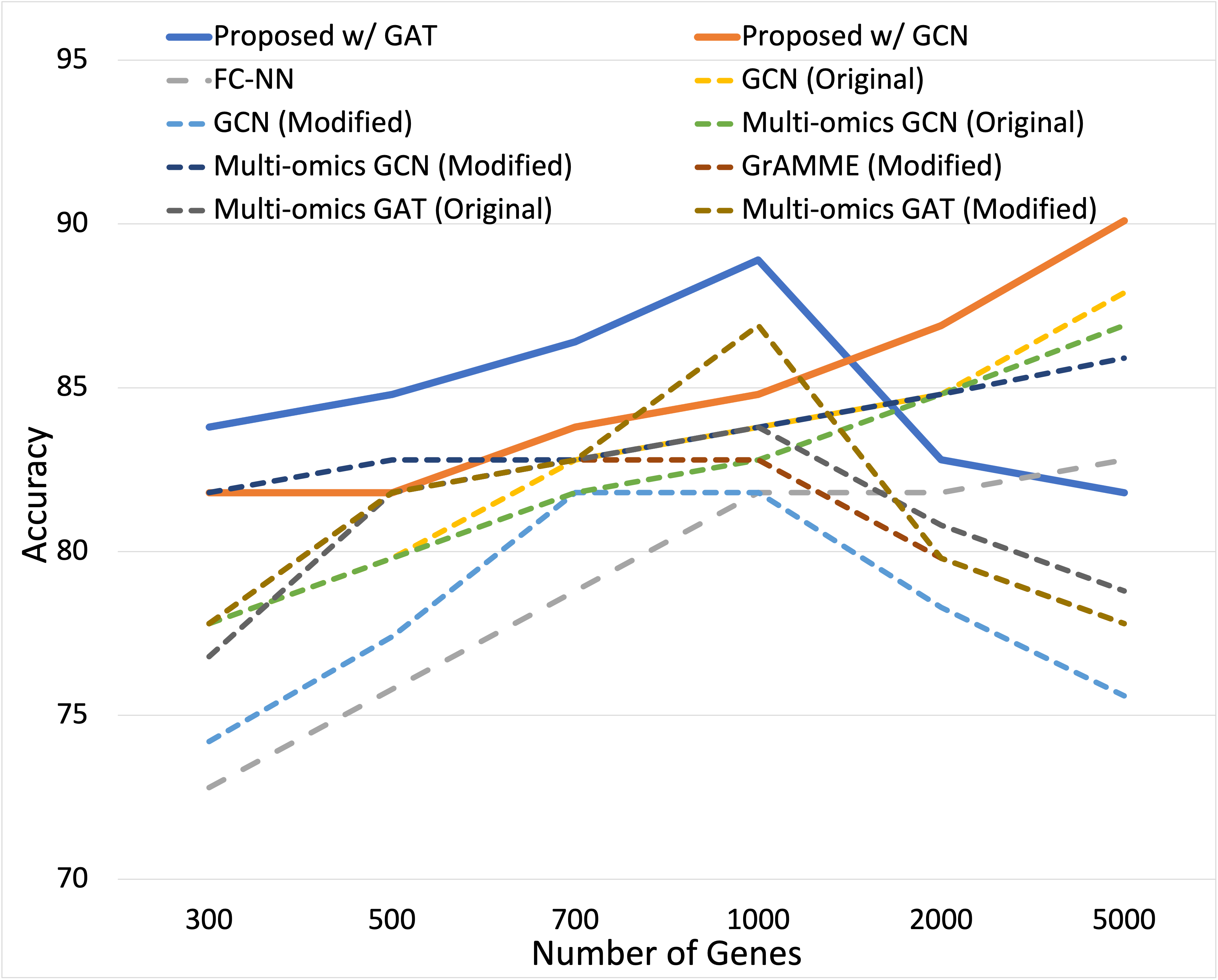}
    \caption{The accuracy of the proposed model with GAT (blue solid line) or GCN (orange solid line) and baseline models (dashed line) are plotted against different numbers of genes (300, 500, 700, 1000, 2000, and 5000) for BRCA subtype classification.}
    \label{fig:brca_1}
\end{subfigure}
\begin{subfigure}{0.9\textwidth}
\centering
   \includegraphics[width=0.95\textwidth]{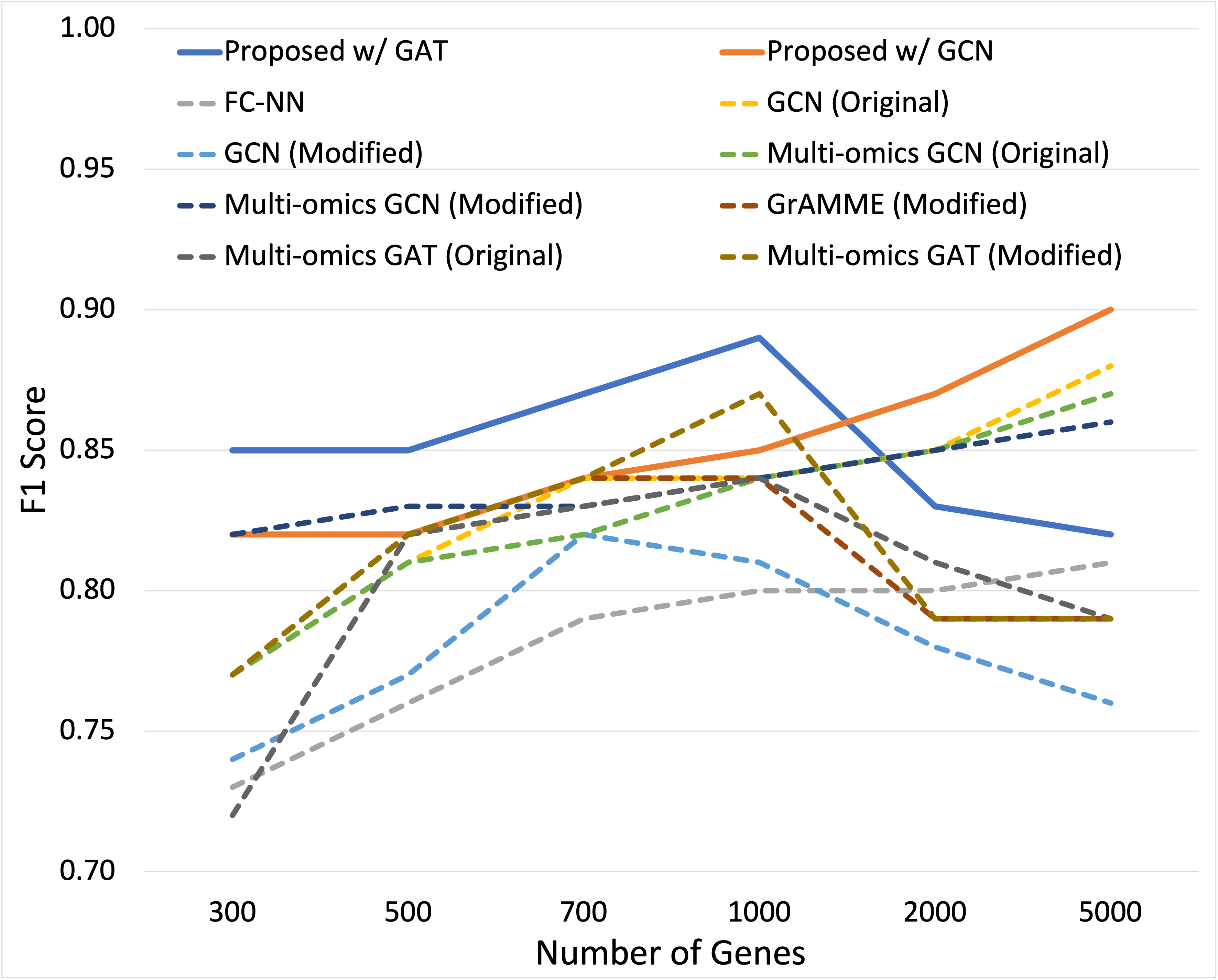}
    \caption{The F1 scores of the proposed model with GAT (blue solid line) or GCN (orange solid line) and baseline models (dashed line) are plotted against different numbers of genes (300, 500, 700, 1000, 2000, and 5000) for BRCA subtype classification.}
    \label{fig:brca_2}
\end{subfigure}
\end{figure*}

The accuracy and F1 scores of the proposed model and the baseline models for BRCA subtype classification are shown in Figure \ref{fig_brca_result}. The proposed model with GAT performs best when the number of genes is smaller than 1000 and the proposed model with GCN performs best when the number of genes is larger than 1000. The proposed GAT-based model yields the best result with an accuracy of 88.9\% and an F1 score of 0.89 when using 700 genes; and the proposed GCN-based model yields the best result with an accuracy of 90.1\% and an F1 score of 0.90 when using 5000 genes. The detailed results are shown in the supplementary file.
The performance of the proposed model with GAT deteriorates beyond 1,000 genes, but the performance of the proposed model with GCN continues to rise as the number of genes grows beyond 1,000 genes. All GAT-based baseline models show similar deterioration around 1000 genes. We think the high computation cost of the GAT-based model can cause it to perform worse on a large graph than on a small graph. Overall, we can conclude that the proposed model with GCN layers scales better than that with GAT layers at a large number of genes. 

In the process of testing the models on a large graph, we also find that a GAT-based model is more stable on a smaller learning rate compared to a GCN-based model. We believe it is caused by GAT's high computation costs since a high learning rate may cause the model to be stuck in a local optimum.

Overall, we see the proposed model achieves the best performance and scales well with a larger number of genes. We can also conclude that more genes and more omics mostly improve the performance of models, the models with more modules have better performance, and GAT-based models perform better with smaller graphs while GCN-based models scale better at larger graphs.

\subsubsection*{Different Training Set Split}
To examine the performance of the proposed model on a complex dataset with a smaller training set, we tested the model on the Pan-cancer dataset using three different training set splits. This approach was taken to mimic situations where only a smaller labeled dataset is available in the real world. The training set splits were set at $70\%$, $60\%$, and $50\%$, with corresponding testing set splits of $20\%$, $30\%$, and $40\%$. Throughout these tests, the validation set split was consistently kept at $10\%$.

\begin{table}[ht]
\centering
\caption{Proposed Model with Different Training-validation-testing Split}
\resizebox{\textwidth}{!}{
\begin{threeparttable}
\begin{tabular}{l|cccccc}
\hline
\multirow{3}{*}{Model} & \multicolumn{6}{c}{Training Set Ratio}                                         \\ \cline{2-7} 
                       & \multicolumn{2}{c}{70\%} & \multicolumn{2}{c}{60\%} & \multicolumn{2}{c}{50\%} \\
                       & Accu.\tnote{1} & F1 & Accu.\tnote{1} & F1 & Accu.\tnote{1} & F1 \\ \hline
Proposed w/ GAT        & $82.5\% \pm 1.5\%$ & $0.82 \pm 0.02$ & $79.9\% \pm 4.0\%$ & $0.78 \pm 0.06$ & $74.2\% \pm 7.5\%$ & $0.71 \pm 0.10$ \\
Proposed w/ GCN        & $77.9\% \pm 1.2\%$ & $0.76 \pm 0.02$ & $76.7\% \pm 0.4\%$ & $0.75 \pm 0.01$ & $77.3\% \pm 2.5\%$ & $0.76 \pm 0.03$ \\ \hline
\multicolumn{7}{l}{\footnotesize\textit{The values after $\pm$ sign are the standard deviations.}}\\
\end{tabular}
\begin{tablenotes}
\item[1] Accu. stands for Accuracy.
\end{tablenotes}
\end{threeparttable}}
\label{tab:train_set}
\end{table}

As shown in Table \ref{tab:train_set}, the proposed model with the GAT layer exhibits a slight performance deterioration at $70\%$ and $60\%$ training set splits. However, it displays a more pronounced decline in classification accuracy at $50\%$. In contrast, the proposed model with the GCN layer demonstrates consistent and robust performance across all three training-validation-testing splits. However, its classification accuracy is lower than that of the model with the GAT layer at $70\%$ and $60\%$ training set splits. Therefore, we can conclude that the proposed model with the GAT layer achieves superior performance compared to the model with the GCN layer when the training set is relatively small. However, the model with the GCN layer outperforms at a very small training set ($50\%$). Overall, the proposed model with the GCN layer offers more robust classification performance with smaller training sets.

\subsubsection*{Different Combinations of Modules}
To examine the effect of different modules within the proposed model, we test three different variants of the proposed model for the Pan-cancer molecular subtype classification. All variants of the proposed model are trained with all three omics data at 300, 500, and 700 genes. The proposed model without the decoder acts as a parallel structured GNN model, the proposed model without the parallel structure acts as a graph autoencoder model, and the proposed model without both the decoder and the parallel structure acts as a graph-classification GNN model.

\begin{table*}[htp]
\centering
\caption{Results of the Variants of the Proposed Model for Molecular Subtype Classification Using the TCGA Pan-cancer Dataset.}
\label{tab:module_comparison}
\resizebox{\textwidth}{!}{
\begin{threeparttable}
\begin{tabular}{l|ll|ll|ll}
\hline
\multirow{2}{*}{GNN Layers (Module)}     & \multicolumn{2}{c|}{300} & \multicolumn{2}{c|}{500} & \multicolumn{2}{c}{700} \\
                                         & Accu.\tnote{1}     & F1       & Accu.\tnote{1}     & F1       & Accu.\tnote{1}     & F1       \\
\hline
GAT (No Decoder)             & \textbf{76.3\%} $\bm{\pm}$ \textbf{1.6\%} & \textbf{0.76} $\bm{\pm}$ \textbf{0.03} &  \textbf{78.2\%} $\bm{\pm}$ \textbf{1.2\%} & \textbf{0.77} $\bm{\pm}$ \textbf{0.01} & \textbf{80.2\%} $\bm{\pm}$ \textbf{1.2\%} & \textbf{0.79} $\bm{\pm}$ \textbf{0.01} \\
GCN (No Decoder)             & 75.3\% $\pm$ 1.2\% & 0.74 $\pm$ 0.02 & 76.8\% $\pm$ 0.8\% &  0.75 $\pm$ 0.01 & 79.3\% $\pm$ 0.8\% & 0.78 $\pm$ 0.01 \\
GAT (No Parallel)            & 75.4\% $\pm$ 1.8\% & 0.73 $\pm$ 0.03 & 76.1\% $\pm$ 1.7\% & 0.73 $\pm$ 0.02 & 79.8\% $\pm$ 1.3\% & 0.78 $\pm$ 0.02 \\
GCN (No Parallel)            & 73.5\% $\pm$ 1.2\% & 0.72 $\pm$ 0.02 &  75.4\% $\pm$ 1.2\%  & 0.73 $\pm$ 0.01 & 76.7\% $\pm$ 0.8\% & 0.75 $\pm$ 0.01 \\
GAT (No Decoder \& Parallel) & 74.9\% $\pm$ 1.4\% & 0.73 $\pm$ 0.02 & 76.4\% $\pm$ 0.9\% & 0.74 $\pm$ 0.01 & 80.1\% $\pm$ 0.8\% & 0.79 $\pm$ 0.01  \\
GCN (No Decoder \& Parallel) & 73.1\% $\pm$ 1.2\% & 0.73 $\pm$ 0.02 & 75.6\% $\pm$ 0.8\% & 0.73 $\pm$ 0.01 & 77.3\% $\pm$ 0.02\% & 0.76 $\pm$ 0.01 \\
\hline
\multicolumn{7}{l}{\footnotesize\textit{The bold font indicates the highest values and the values after $\pm$ sign are the standard deviations.}}\\
\end{tabular}
\begin{tablenotes}
\item[1] Accu. stands for Accuracy.
\end{tablenotes}
\end{threeparttable}
}
\end{table*}

As shown in Table \ref{tab:module_comparison}, models without the parallel structure perform poorly compared to those without the decoder at any number of genes in general. It shows that the parallel structure plays an important role in feature extraction, which also demonstrates the benefit of including both local features and global features. When the graph size is small (300 genes), the model without the decoder and the parallel structure performs more poorly compared to those with either component. However, when the graph size is large enough (500 genes and 700 genes), the model without the decoder and the parallel structure performs relatively the same compared to those with either of the component. We believe the extra information in the large graph compensates for the loss in performance caused by the exclusion of either the decoder or the parallel structure.

\subsubsection*{Different Combination of Omics and Graphs}
\begin{table*}[htp]
\centering
\caption{Results of the Proposed Model on Different Combinations of Omics and Networks at 500 Genes Using the TCGA Pan-cancer Dataset.}
\label{tab:ablation}
\centering
\resizebox{\textwidth}{!}{
\begin{threeparttable}
\begin{tabular}{cc|cc|cc}
\hline
\multirow{2}{*}{Data}           & \multirow{2}{*}{Network} & \multicolumn{2}{c|}{GAT} & \multicolumn{2}{c}{GCN} \\
                                &                          & Accu.\tnote{5}     & F1       & Accu.\tnote{5}      & F1       \\
\hline
mRNA\tnote{1}                  & Intra-omic\tnote{3}  & 77.0\% $\pm$ 1.9\% & 0.75 $\pm$ 0.03 & 76.1\% $\pm$ 0.9\% & 0.73 $\pm$ 0.01 \\
miRNA\tnote{2}                 & Intra-omic\tnote{4}  & 74.0\% $\pm$ 0.4\% & 0.70 $\pm$ 0.01 & 68.2\% $\pm$ 4.1\% & 0.63 $\pm$ 0.04\\
mRNA+CNV\tnote{1}              & Intra-omic\tnote{3}  & 79.1\% $\pm$ 1.4\% & 0.77 $\pm$ 0.03 & 77.1\% $\pm$ 0.7\% & 0.76 $\pm$ 0.01 \\
\multirow{2}{*}{mRNA+miRNA}     & Inter-omic          & 76.1\% $\pm$ 1.6\% & 0.73 $\pm$ 0.03 & 75.4\% $\pm$ 0.7\% & 0.73 $\pm$ 0.01 \\
                                & Intra-omic          & 77.3\% $\pm$ 1.6\% & 0.75 $\pm$ 0.03 & 76.8\% $\pm$ 0.7\% & 0.74 $\pm$ 0.01 \\
\multirow{2}{*}{mRNA+CNV+miRNA} & Inter-omic          & 80.3\% $\pm$ 1.6\% & 0.80 $\pm$ 0.02 & 77.4\% $\pm$ 0.6\% & 0.74 $\pm$ 0.01 \\
                                & Intra-omic          & \textbf{80.5\%} $\bm{\pm}$ \textbf{1.2\%} & \textbf{0.80} $\bm{\pm}$ \textbf{0.02} & \textbf{78.2\%} $\bm{\pm}$ 0.6\% & \textbf{0.75} $\bm{\pm}$ \textbf{0.01} \\
\hline
\multicolumn{6}{l}{\footnotesize\textit{The bold font indicates the highest values and the values after $\pm$ sign are the standard deviations.}}\\
\end{tabular}
\begin{tablenotes}
\item[1] Data contains no miRNA-based nodes, so only 500 gene nodes in the graph
\item[2] Data contains no gene-based nodes, so only 100 miRNA nodes in the graph
\item[3] The graph contains only gene-gene connections.
\item[4] The graph contains only miRNA-miRNA meta-path connections.
\item[5] Accu. stands for accuracy.
\end{tablenotes}
\end{threeparttable}
}
\end{table*}

To test the effect of different choices of omics and different graphs, we generate five different combinations of omics. The five combinations of omics are mRNA, miRNA, mRNA + CNV, mRNA + miRNA, and mRNA + CNV + miRNA. For mRNA + miRNA and mRNA + CNV + miRNA, two different variants of graphs are also tested. All models are conducted for Pan-cancer molecular subtype classification, and trained with 500 genes except for only miRNA omic, which contains only 100 miRNA nodes. As shown in Table \ref{tab:ablation}, the best-performing setting is mRNA + CNV + miRNA with intra-omic edges for both GAT-based and GCN-based models. The worst-performing setting is miRNA, which has the smallest graph size and information. Models on mRNA + CNV perform better than those on mRNA + miRNA, but adding miRNA to mRNA + CNV (mRNA + CNV + miRNA setting) still improves the model performance. Models with intra-omic graph performs slightly better than models with inter-omics graph. The performance difference across different settings is the same for both GAT-based and GCN-based models.

\section*{Conclusion}
In this study, we propose a novel end-to-end multi-omics GNN framework for accurate and robust cancer subtype classification. The proposed model utilizes multi-omics data in the form of a heterogeneous multi-layer graph, which is the supra-graph built from GGI network, miRNA-gene target network, and miRNA meta-path. While GNNs have been previously employed for genomics data analysis, our model's novelty lies in the utilization of a heterogeneous multi-layer multiomics supra-graph. The supra-graph not only incorporates inter-omics and intra-omic connections from established biological knowledge but also integrates genomics, transcriptomics, and epigenomics data into a single graph, providing a novel advancement in cancer subtype classification. The proposed model outperforms all four baseline models for cancer molecular subtype classification. We do a thorough comparative analysis of GAT and GCN-based models at different numbers of gene settings, different combinations of omics, and different graphs.

Comparing the proposed model to the baseline models, it achieves the best performance for cancer molecular subtype classification and BRCA subtype classification. The proposed model with GAT layers performs better than that with GCN layers at smaller-size graphs (smaller than 1,000 genes). However, the performance of the GAT-based model deteriorates as the size of the graph grows beyond a certain threshold. On the other hand, the performance of the GCN-based model continues to improve as the size of the graph grows. Therefore, we can conclude that a GAT-based model is more suitable on a smaller graph, where it has a higher feature extraction ability and its computation cost isn't that high yet.

By studying the effect of different modules within the proposed model and different combinations of omics, we find the addition of a decoder and the parallel structure, and including other omics improves the performance of the proposed model. The benefit of using parallel structure outweighs that of decoder, especially on smaller-size graphs, and the benefit of adding CNV is higher than that of adding miRNA. We also find that using a graph with only intra-omic edges yields a better performance than using a graph with only inter-omics edges, which agrees with the results from the previous study~\cite{Kaczmarek2022}.

The proposed model also has some limitations. We investigate only two well-established and widely adopted GNN models. New models are emerging with the recent blooming of studies in GNN models. As the size of the graph grows or more omics are added, GAT-based models become more sensitive to parameters and take a much longer time to train. It is our future research direction to overcome such limitations. The proposed model for cancer subtype classification depends on labeled data, which is costly to annotate and difficult to obtain in the real world. Exploring unsupervised learning for cancer subtype detection is also a direction we aim to pursue in our future research.

In summary, incorporating gene-based and non-gene-based omic data in the form of a supra-graph with inter-omics and intra-omic connections improves the cancer subtype classification. The GAT-based model is preferable on smaller graphs than the GCN-based model. GCN-based model is preferable when dealing with large and complex graphs.



\section*{Declarations}
\begin{backmatter}
\section*{Acknowledgements}
Not applicable.

\section*{Funding}
This work is supported by the National Science Foundation (NSF) under grant No. 1942303, PI: Nabavi.

\section*{Abbreviations}
Not applicable.

\section*{Availability of data and materials}
TCGA Pan-cancer dataset and TCGA BRCA dataset are both obtained from Xena database (https://xenabrowser.net), the detailed link for TCGA Pan-cancer dataset is (\url{https://xenabrowser.net/datapages/?cohort=TCGA%20Pan-Cancer%20(PANCAN)&removeHub=https%3A%2F%2Fxena.treehouse.gi.ucsc.edu%3A443}) and the detailed link for TCGA BRCA dataset is (\url{https://xenabrowser.net/datapages/?cohort=TCGA%20Breast%20Cancer%20(BRCA)&removeHub=https%3A%2F%2Fxena.treehouse.gi.ucsc.edu%3A443})~\cite{goldman2020visualizing}. The code for the proposed method can be found at our Github repository (https://github.com/NabaviLab/Multimodal-GNN-for-Cancer-Subtype-Clasification).

\section*{Ethics approval and consent to participate}
Not applicable.

\section*{Competing interests}
The authors declare that they have no competing interests.

\section*{Consent for publication}
Not applicable.

\section*{Authors' contributions}
B.L. obtained the TCGA data and network data. B.L. designed the new method and analyzed the results. B.L. and S.N. drafted the manuscript and revised the manuscript together. Both authors have approved the final manuscript.



\bibliographystyle{bmc-mathphys} 
\bibliography{bmc_article}      


\begin{thebibliography}{34}
\ifx \bisbn   \undefined \def \bisbn  #1{ISBN #1}\fi
\ifx \binits  \undefined \def \binits#1{#1}\fi
\ifx \bauthor  \undefined \def \bauthor#1{#1}\fi
\ifx \batitle  \undefined \def \batitle#1{#1}\fi
\ifx \bjtitle  \undefined \def \bjtitle#1{#1}\fi
\ifx \bvolume  \undefined \def \bvolume#1{\textbf{#1}}\fi
\ifx \byear  \undefined \def \byear#1{#1}\fi
\ifx \bissue  \undefined \def \bissue#1{#1}\fi
\ifx \bfpage  \undefined \def \bfpage#1{#1}\fi
\ifx \blpage  \undefined \def \blpage #1{#1}\fi
\ifx \burl  \undefined \def \burl#1{\textsf{#1}}\fi
\ifx \doiurl  \undefined \def \doiurl#1{\textsf{#1}}\fi
\ifx \betal  \undefined \def \betal{\textit{et al.}}\fi
\ifx \binstitute  \undefined \def \binstitute#1{#1}\fi
\ifx \binstitutionaled  \undefined \def \binstitutionaled#1{#1}\fi
\ifx \bctitle  \undefined \def \bctitle#1{#1}\fi
\ifx \beditor  \undefined \def \beditor#1{#1}\fi
\ifx \bpublisher  \undefined \def \bpublisher#1{#1}\fi
\ifx \bbtitle  \undefined \def \bbtitle#1{#1}\fi
\ifx \bedition  \undefined \def \bedition#1{#1}\fi
\ifx \bseriesno  \undefined \def \bseriesno#1{#1}\fi
\ifx \blocation  \undefined \def \blocation#1{#1}\fi
\ifx \bsertitle  \undefined \def \bsertitle#1{#1}\fi
\ifx \bsnm \undefined \def \bsnm#1{#1}\fi
\ifx \bsuffix \undefined \def \bsuffix#1{#1}\fi
\ifx \bparticle \undefined \def \bparticle#1{#1}\fi
\ifx \barticle \undefined \def \barticle#1{#1}\fi
\ifx \bconfdate \undefined \def \bconfdate #1{#1}\fi
\ifx \botherref \undefined \def \botherref #1{#1}\fi
\ifx \url \undefined \def \url#1{\textsf{#1}}\fi
\ifx \bchapter \undefined \def \bchapter#1{#1}\fi
\ifx \bbook \undefined \def \bbook#1{#1}\fi
\ifx \bcomment \undefined \def \bcomment#1{#1}\fi
\ifx \oauthor \undefined \def \oauthor#1{#1}\fi
\ifx \citeauthoryear \undefined \def \citeauthoryear#1{#1}\fi
\ifx \endbibitem  \undefined \def \endbibitem {}\fi
\ifx \bconflocation  \undefined \def \bconflocation#1{#1}\fi
\ifx \arxivurl  \undefined \def \arxivurl#1{\textsf{#1}}\fi
\csname PreBibitemsHook\endcsname

\bibitem{Li2021}
\begin{botherref}
\oauthor{\bsnm{Li}, \binits{B.}},
\oauthor{\bsnm{Wang}, \binits{T.}},
\oauthor{\bsnm{Nabavi}, \binits{S.}}:
Cancer molecular subtype classification by graph convolutional networks on
  multi-omics data.
Proceedings of the 12th ACM Conference on Bioinformatics, Computational
  Biology, and Health Informatics, BCB 2021
\textbf{1}
(2021).
doi:\doiurl{10.1145/3459930.3469542}
\end{botherref}
\endbibitem

\bibitem{Zhang2019}
\begin{botherref}
\oauthor{\bsnm{Zhang}, \binits{X.}},
\oauthor{\bsnm{Zhang}, \binits{J.}},
\oauthor{\bsnm{Sun}, \binits{K.}},
\oauthor{\bsnm{Yang}, \binits{X.}},
\oauthor{\bsnm{Dai}, \binits{C.}},
\oauthor{\bsnm{Guo}, \binits{Y.}}:
Integrated multi-omics analysis using variational autoencoders: Application to
  pan-cancer classification.
Proceedings - 2019 IEEE International Conference on Bioinformatics and
  Biomedicine, BIBM 2019,
765--769
(2019).
doi:\doiurl{10.1109/BIBM47256.2019.8983228}
\end{botherref}
\endbibitem

\bibitem{Yang2019}
\begin{barticle}
\bauthor{\bsnm{Yang}, \binits{B.}},
\bauthor{\bsnm{Zhang}, \binits{Y.}},
\bauthor{\bsnm{Pang}, \binits{S.}},
\bauthor{\bsnm{Shang}, \binits{X.}},
\bauthor{\bsnm{Zhao}, \binits{X.}},
\bauthor{\bsnm{Han}, \binits{M.}}:
\batitle{Integrating multi-omic data with deep subspace fusion clustering for
  cancer subtype prediction}.
\bjtitle{IEEE/ACM Transactions on Computational Biology and Bioinformatics}
\bvolume{XX},
\bfpage{1}--\blpage{1}
(\byear{2019}).
doi:\doiurl{10.1109/tcbb.2019.2951413}
\end{barticle}
\endbibitem

\bibitem{sharifi2019}
\begin{barticle}
\bauthor{\bsnm{Sharifi-Noghabi}, \binits{H.}},
\bauthor{\bsnm{Zolotareva}, \binits{O.}},
\bauthor{\bsnm{Collins}, \binits{C.C.}},
\bauthor{\bsnm{Ester}, \binits{M.}}:
\batitle{Moli: Multi-omics late integration with deep neural networks for drug
  response prediction}.
\bjtitle{Bioinformatics}
\bvolume{35},
\bfpage{501}--\blpage{509}
(\byear{2019}).
doi:\doiurl{10.1093/bioinformatics/btz318}
\end{barticle}
\endbibitem

\bibitem{Wang2021}
\begin{barticle}
\bauthor{\bsnm{Wang}, \binits{T.}},
\bauthor{\bsnm{Shao}, \binits{W.}},
\bauthor{\bsnm{Huang}, \binits{Z.}},
\bauthor{\bsnm{Tang}, \binits{H.}},
\bauthor{\bsnm{Zhang}, \binits{J.}},
\bauthor{\bsnm{Ding}, \binits{Z.}},
\bauthor{\bsnm{Huang}, \binits{K.}}:
\batitle{Mogonet integrates multi-omics data using graph convolutional networks
  allowing patient classification and biomarker identification}.
\bjtitle{Nature Communications}
\bvolume{12},
\bfpage{3445}
(\byear{2021}).
doi:\doiurl{10.1038/s41467-021-23774-w}
\end{barticle}
\endbibitem

\bibitem{Ma2019}
\begin{barticle}
\bauthor{\bsnm{Ma}, \binits{T.}},
\bauthor{\bsnm{Zhang}, \binits{A.}}:
\batitle{Integrate multi-omics data with biological interaction networks using
  multi-view factorization autoencoder (mae)}.
\bjtitle{BMC Genomics}
\bvolume{20},
\bfpage{1}--\blpage{11}
(\byear{2019}).
doi:\doiurl{10.1186/s12864-019-6285-x}
\end{barticle}
\endbibitem

\bibitem{Kaczmarek2022}
\begin{barticle}
\bauthor{\bsnm{Kaczmarek}, \binits{E.}},
\bauthor{\bsnm{Jamzad}, \binits{A.}},
\bauthor{\bsnm{Imtiaz}, \binits{T.}},
\bauthor{\bsnm{Nanayakkara}, \binits{J.}},
\bauthor{\bsnm{Renwick}, \binits{N.}},
\bauthor{\bsnm{Mousavi}, \binits{P.}}:
\batitle{Multi-omic graph transformers for cancer classification and
  interpretation}.
\bjtitle{Pacific Symposium on Biocomputing. Pacific Symposium on Biocomputing}
\bvolume{27},
\bfpage{373}--\blpage{384}
(\byear{2022})
\end{barticle}
\endbibitem

\bibitem{Lotfollahi2022}
\begin{botherref}
\oauthor{\bsnm{Lotfollahi}, \binits{M.}},
\oauthor{\bsnm{Litinetskaya}, \binits{A.}},
\oauthor{\bsnm{Theis}, \binits{F.J.}}:
Multigrate : single-cell multi-omic data integration,
1--5
(2022).
doi:\doiurl{10.1101/2022.03.16.484643}
\end{botherref}
\endbibitem

\bibitem{Huang2019}
\begin{barticle}
\bauthor{\bsnm{Huang}, \binits{Z.}},
\bauthor{\bsnm{Zhan}, \binits{X.}},
\bauthor{\bsnm{Xiang}, \binits{S.}},
\bauthor{\bsnm{Johnson}, \binits{T.S.}},
\bauthor{\bsnm{Helm}, \binits{B.}},
\bauthor{\bsnm{Yu}, \binits{C.Y.}},
\bauthor{\bsnm{Zhang}, \binits{J.}},
\bauthor{\bsnm{Salama}, \binits{P.}},
\bauthor{\bsnm{Rizkalla}, \binits{M.}},
\bauthor{\bsnm{Han}, \binits{Z.}},
\bauthor{\bsnm{Huang}, \binits{K.}}:
\batitle{Salmon: Survival analysis learning with multi-omics neural networks on
  breast cancer}.
\bjtitle{Frontiers in Genetics}
\bvolume{10},
\bfpage{1}--\blpage{13}
(\byear{2019}).
doi:\doiurl{10.3389/fgene.2019.00166}
\end{barticle}
\endbibitem

\bibitem{bai2022semi}
\begin{bchapter}
\bauthor{\bsnm{Bai}, \binits{J.}},
\bauthor{\bsnm{Li}, \binits{B.}},
\bauthor{\bsnm{Nabavi}, \binits{S.}}:
\bctitle{Semi-supervised classification of disease prognosis using cr images
  with clinical data structured graph}.
In: \bbtitle{Proceedings of the 13th ACM International Conference on
  Bioinformatics, Computational Biology and Health Informatics},
pp. \bfpage{1}--\blpage{9}
(\byear{2022})
\end{bchapter}
\endbibitem

\bibitem{chai2021integrating}
\begin{barticle}
\bauthor{\bsnm{Chai}, \binits{H.}},
\bauthor{\bsnm{Zhou}, \binits{X.}},
\bauthor{\bsnm{Zhang}, \binits{Z.}},
\bauthor{\bsnm{Rao}, \binits{J.}},
\bauthor{\bsnm{Zhao}, \binits{H.}},
\bauthor{\bsnm{Yang}, \binits{Y.}}:
\batitle{Integrating multi-omics data through deep learning for accurate cancer
  prognosis prediction}.
\bjtitle{Computers in biology and medicine}
\bvolume{134},
\bfpage{104481}
(\byear{2021})
\end{barticle}
\endbibitem

\bibitem{Heo2021}
\begin{barticle}
\bauthor{\bsnm{Heo}, \binits{Y.J.}},
\bauthor{\bsnm{Hwa}, \binits{C.}},
\bauthor{\bsnm{Lee}, \binits{G.H.}},
\bauthor{\bsnm{Park}, \binits{J.M.}},
\bauthor{\bsnm{An}, \binits{J.Y.}}:
\batitle{Integrative multi-omics approaches in cancer research: From biological
  networks to clinical subtypes}.
\bjtitle{Molecules and Cells}
\bvolume{44},
\bfpage{433}--\blpage{443}
(\byear{2021}).
doi:\doiurl{10.14348/molcells.2021.0042}
\end{barticle}
\endbibitem

\bibitem{hoadley2014multiplatform}
\begin{barticle}
\bauthor{\bsnm{Hoadley}, \binits{K.A.}},
\bauthor{\bsnm{Yau}, \binits{C.}},
\bauthor{\bsnm{Wolf}, \binits{D.M.}},
\bauthor{\bsnm{Cherniack}, \binits{A.D.}},
\bauthor{\bsnm{Tamborero}, \binits{D.}},
\bauthor{\bsnm{Ng}, \binits{S.}},
\bauthor{\bsnm{Leiserson}, \binits{M.D.}},
\bauthor{\bsnm{Niu}, \binits{B.}},
\bauthor{\bsnm{McLellan}, \binits{M.D.}},
\bauthor{\bsnm{Uzunangelov}, \binits{V.}}, \betal:
\batitle{Multiplatform analysis of 12 cancer types reveals molecular
  classification within and across tissues of origin}.
\bjtitle{Cell}
\bvolume{158}(\bissue{4}),
\bfpage{929}--\blpage{944}
(\byear{2014})
\end{barticle}
\endbibitem

\bibitem{mateo2022delivering}
\begin{barticle}
\bauthor{\bsnm{Mateo}, \binits{J.}},
\bauthor{\bsnm{Steuten}, \binits{L.}},
\bauthor{\bsnm{Aftimos}, \binits{P.}},
\bauthor{\bsnm{Andr{\'e}}, \binits{F.}},
\bauthor{\bsnm{Davies}, \binits{M.}},
\bauthor{\bsnm{Garralda}, \binits{E.}},
\bauthor{\bsnm{Geissler}, \binits{J.}},
\bauthor{\bsnm{Husereau}, \binits{D.}},
\bauthor{\bsnm{Martinez-Lopez}, \binits{I.}},
\bauthor{\bsnm{Normanno}, \binits{N.}}, \betal:
\batitle{Delivering precision oncology to patients with cancer}.
\bjtitle{Nature Medicine}
\bvolume{28}(\bissue{4}),
\bfpage{658}--\blpage{665}
(\byear{2022})
\end{barticle}
\endbibitem

\bibitem{hoadley2018cell}
\begin{barticle}
\bauthor{\bsnm{Hoadley}, \binits{K.A.}},
\bauthor{\bsnm{Yau}, \binits{C.}},
\bauthor{\bsnm{Hinoue}, \binits{T.}},
\bauthor{\bsnm{Wolf}, \binits{D.M.}},
\bauthor{\bsnm{Lazar}, \binits{A.J.}},
\bauthor{\bsnm{Drill}, \binits{E.}},
\bauthor{\bsnm{Shen}, \binits{R.}},
\bauthor{\bsnm{Taylor}, \binits{A.M.}},
\bauthor{\bsnm{Cherniack}, \binits{A.D.}},
\bauthor{\bsnm{Thorsson}, \binits{V.}}, \betal:
\batitle{Cell-of-origin patterns dominate the molecular classification of
  10,000 tumors from 33 types of cancer}.
\bjtitle{Cell}
\bvolume{173}(\bissue{2}),
\bfpage{291}--\blpage{304}
(\byear{2018})
\end{barticle}
\endbibitem

\bibitem{Defferrard2016}
\begin{botherref}
\oauthor{\bsnm{Defferrard}, \binits{M.}},
\oauthor{\bsnm{Bresson}, \binits{X.}},
\oauthor{\bsnm{Vandergheynst}, \binits{P.}}:
Convolutional neural networks on graphs with fast localized spectral filtering.
Advances in Neural Information Processing Systems,
3844--3852
(2016)
\end{botherref}
\endbibitem

\bibitem{zou2019primer}
\begin{barticle}
\bauthor{\bsnm{Zou}, \binits{J.}},
\bauthor{\bsnm{Huss}, \binits{M.}},
\bauthor{\bsnm{Abid}, \binits{A.}},
\bauthor{\bsnm{Mohammadi}, \binits{P.}},
\bauthor{\bsnm{Torkamani}, \binits{A.}},
\bauthor{\bsnm{Telenti}, \binits{A.}}:
\batitle{A primer on deep learning in genomics}.
\bjtitle{Nature genetics}
\bvolume{51}(\bissue{1}),
\bfpage{12}--\blpage{18}
(\byear{2019})
\end{barticle}
\endbibitem

\bibitem{he2020data}
\begin{botherref}
\oauthor{\bsnm{He}, \binits{S.}},
\oauthor{\bsnm{Pepin}, \binits{L.}},
\oauthor{\bsnm{Wang}, \binits{G.}},
\oauthor{\bsnm{Zhang}, \binits{D.}},
\oauthor{\bsnm{Miao}, \binits{F.}}:
Data-driven distributionally robust electric vehicle balancing for
  mobility-on-demand systems under demand and supply uncertainties.
In: 2020 IEEE/RSJ International Conference on Intelligent Robots and Systems
  (IROS),
pp. 2165--2172.
IEEE
\end{botherref}
\endbibitem

\bibitem{wang2021clustering}
\begin{bchapter}
\bauthor{\bsnm{Wang}, \binits{T.}},
\bauthor{\bsnm{Li}, \binits{B.}},
\bauthor{\bsnm{Nabavi}, \binits{S.}}:
\bctitle{Single-cell rna sequencing data clustering using graph convolutional
  networks}.
In: \bbtitle{2021 IEEE International Conference on Bioinformatics and
  Biomedicine (BIBM)},
pp. \bfpage{2163}--\blpage{2170}
(\byear{2021}).
\bcomment{IEEE}
\end{bchapter}
\endbibitem

\bibitem{shi2022constraint}
\begin{botherref}
\oauthor{\bsnm{Shi}, \binits{C.}},
\oauthor{\bsnm{Emadikhiav}, \binits{M.}},
\oauthor{\bsnm{Lozano}, \binits{L.}},
\oauthor{\bsnm{Bergman}, \binits{D.}}:
Constraint learning to define trust regions in predictive-model embedded
  optimization.
arXiv preprint arXiv:2201.04429
(2022)
\end{botherref}
\endbibitem

\bibitem{he2023robust}
\begin{bchapter}
\bauthor{\bsnm{He}, \binits{S.}},
\bauthor{\bsnm{Han}, \binits{S.}},
\bauthor{\bsnm{Miao}, \binits{F.}}:
\bctitle{Robust electric vehicle balancing of autonomous mobility-on-demand
  system: A multi-agent reinforcement learning approach}.
In: \bbtitle{2023 IEEE/RSJ International Conference on Intelligent Robots and
  Systems (IROS)},
pp. \bfpage{5471}--\blpage{5478}
(\byear{2023}).
\bcomment{IEEE}
\end{bchapter}
\endbibitem

\bibitem{wang2021two}
\begin{bchapter}
\bauthor{\bsnm{Wang}, \binits{K.}},
\bauthor{\bsnm{Lozano}, \binits{L.}},
\bauthor{\bsnm{Bergman}, \binits{D.}},
\bauthor{\bsnm{Cardonha}, \binits{C.}}:
\bctitle{A two-stage exact algorithm for optimization of neural network
  ensemble}.
In: \bbtitle{Integration of Constraint Programming, Artificial Intelligence,
  and Operations Research: 18th International Conference, CPAIOR 2021, Vienna,
  Austria, July 5--8, 2021, Proceedings 18},
pp. \bfpage{106}--\blpage{114}
(\byear{2021}).
\bcomment{Springer}
\end{bchapter}
\endbibitem

\bibitem{nicora2020integrated}
\begin{barticle}
\bauthor{\bsnm{Nicora}, \binits{G.}},
\bauthor{\bsnm{Vitali}, \binits{F.}},
\bauthor{\bsnm{Dagliati}, \binits{A.}},
\bauthor{\bsnm{Geifman}, \binits{N.}},
\bauthor{\bsnm{Bellazzi}, \binits{R.}}:
\batitle{Integrated multi-omics analyses in oncology: a review of machine
  learning methods and tools}.
\bjtitle{Frontiers in oncology}
\bvolume{10},
\bfpage{1030}
(\byear{2020})
\end{barticle}
\endbibitem

\bibitem{wu2020comprehensive}
\begin{botherref}
\oauthor{\bsnm{Wu}, \binits{Z.}},
\oauthor{\bsnm{Pan}, \binits{S.}},
\oauthor{\bsnm{Chen}, \binits{F.}},
\oauthor{\bsnm{Long}, \binits{G.}},
\oauthor{\bsnm{Zhang}, \binits{C.}},
\oauthor{\bsnm{Philip}, \binits{S.Y.}}:
A comprehensive survey on graph neural networks.
IEEE transactions on neural networks and learning systems
(2020)
\end{botherref}
\endbibitem

\bibitem{petar2017}
\begin{botherref}
\oauthor{\bsnm{Velicković}, \binits{P.}},
\oauthor{\bsnm{Cucurull}, \binits{G.}},
\oauthor{\bsnm{Casanova}, \binits{A.}},
\oauthor{\bsnm{Romero}, \binits{A.}},
\oauthor{\bsnm{Liò}, \binits{P.}},
\oauthor{\bsnm{Bengio}, \binits{Y.}}:
Graph attention networks.
arXiv,
1--12
(2017)
\end{botherref}
\endbibitem

\bibitem{ramirez2020classification}
\begin{botherref}
\oauthor{\bsnm{Ramirez}, \binits{R.}},
\oauthor{\bsnm{Chiu}, \binits{Y.-C.}},
\oauthor{\bsnm{Hererra}, \binits{A.}},
\oauthor{\bsnm{Mostavi}, \binits{M.}},
\oauthor{\bsnm{Ramirez}, \binits{J.}},
\oauthor{\bsnm{Chen}, \binits{Y.}},
\oauthor{\bsnm{Huang}, \binits{Y.}},
\oauthor{\bsnm{Jin}, \binits{Y.-F.}}:
Classification of cancer types using graph convolutional neural networks.
Frontiers in physics
\textbf{8}
(2020)
\end{botherref}
\endbibitem

\bibitem{wang2021single}
\begin{barticle}
\bauthor{\bsnm{Wang}, \binits{T.}},
\bauthor{\bsnm{Bai}, \binits{J.}},
\bauthor{\bsnm{Nabavi}, \binits{S.}}:
\batitle{Single-cell classification using graph convolutional networks}.
\bjtitle{BMC bioinformatics}
\bvolume{22}(\bissue{1}),
\bfpage{1}--\blpage{23}
(\byear{2021})
\end{barticle}
\endbibitem

\bibitem{Shanthamallu2020}
\begin{barticle}
\bauthor{\bsnm{Shanthamallu}, \binits{U.S.}},
\bauthor{\bsnm{Thiagarajan}, \binits{J.J.}},
\bauthor{\bsnm{Song}, \binits{H.}},
\bauthor{\bsnm{Spanias}, \binits{A.}}:
\batitle{Gramme: Semisupervised learning using multilayered graph attention
  models}.
\bjtitle{IEEE Transactions on Neural Networks and Learning Systems}
\bvolume{31},
\bfpage{3977}--\blpage{3988}
(\byear{2020}).
doi:\doiurl{10.1109/TNNLS.2019.2948797}
\end{barticle}
\endbibitem

\bibitem{onitilo2009breast}
\begin{barticle}
\bauthor{\bsnm{Onitilo}, \binits{A.A.}},
\bauthor{\bsnm{Engel}, \binits{J.M.}},
\bauthor{\bsnm{Greenlee}, \binits{R.T.}},
\bauthor{\bsnm{Mukesh}, \binits{B.N.}}:
\batitle{Breast cancer subtypes based on er/pr and her2 expression: comparison
  of clinicopathologic features and survival}.
\bjtitle{Clinical medicine \& research}
\bvolume{7}(\bissue{1-2}),
\bfpage{4}--\blpage{13}
(\byear{2009})
\end{barticle}
\endbibitem

\bibitem{pmid33070389}
\begin{barticle}
\bauthor{\bsnm{Oughtred}, \binits{R.}},
\bauthor{\bsnm{Rust}, \binits{J.}},
\bauthor{\bsnm{Chang}, \binits{C.}},
\bauthor{\bsnm{Breitkreutz}, \binits{B.J.}},
\bauthor{\bsnm{Stark}, \binits{C.}},
\bauthor{\bsnm{Willems}, \binits{A.}},
\bauthor{\bsnm{Boucher}, \binits{L.}},
\bauthor{\bsnm{Leung}, \binits{G.}},
\bauthor{\bsnm{Kolas}, \binits{N.}},
\bauthor{\bsnm{Zhang}, \binits{F.}},
\bauthor{\bsnm{Dolma}, \binits{S.}},
\bauthor{\bsnm{Coulombe-Huntington}, \binits{J.}},
\bauthor{\bsnm{Chatr-Aryamontri}, \binits{A.}},
\bauthor{\bsnm{Dolinski}, \binits{K.}},
\bauthor{\bsnm{Tyers}, \binits{M.}}:
\batitle{{{T}he {B}io{G}{R}{I}{D} database: {A} comprehensive biomedical
  resource of curated protein, genetic, and chemical interactions}}.
\bjtitle{Protein Sci}
\bvolume{30}(\bissue{1}),
\bfpage{187}--\blpage{200}
(\byear{2021})
\end{barticle}
\endbibitem

\bibitem{chen2020mirdb}
\begin{barticle}
\bauthor{\bsnm{Chen}, \binits{Y.}},
\bauthor{\bsnm{Wang}, \binits{X.}}:
\batitle{mirdb: an online database for prediction of functional microrna
  targets}.
\bjtitle{Nucleic acids research}
\bvolume{48}(\bissue{D1}),
\bfpage{127}--\blpage{131}
(\byear{2020})
\end{barticle}
\endbibitem

\bibitem{Lee2020}
\begin{barticle}
\bauthor{\bsnm{Lee}, \binits{B.}},
\bauthor{\bsnm{Zhang}, \binits{S.}},
\bauthor{\bsnm{Poleksic}, \binits{A.}},
\bauthor{\bsnm{Xie}, \binits{L.}}:
\batitle{Heterogeneous multi-layered network model for omics data integration
  and analysis}.
\bjtitle{Frontiers in Genetics}
\bvolume{10},
\bfpage{1}--\blpage{11}
(\byear{2020}).
doi:\doiurl{10.3389/fgene.2019.01381}
\end{barticle}
\endbibitem

\bibitem{cancer2012comprehensive}
\begin{barticle}
\bauthor{\bsnm{13}, \binits{B..W.H..H.M.S.C.L...P.P.J..K.R.}},
\bauthor{\bparticle{data analysis: Baylor College~of} \bsnm{Medicine Creighton
  Chad J. 22 23 Donehower Lawrence A. 22 23 24~25}, \binits{G.}},
\bauthor{\bparticle{for} \bsnm{Systems Biology Reynolds Sheila 31 Kreisberg
  Richard B. 31 Bernard Brady 31 Bressler Ryan 31 Erkkila Timo 32 Lin Jake 31
  Thorsson Vesteinn 31 Zhang Wei 33 Shmulevich Ilya~31}, \binits{I.}}, \betal:
\batitle{Comprehensive molecular portraits of human breast tumours}.
\bjtitle{Nature}
\bvolume{490}(\bissue{7418}),
\bfpage{61}--\blpage{70}
(\byear{2012})
\end{barticle}
\endbibitem

\bibitem{goldman2020visualizing}
\begin{barticle}
\bauthor{\bsnm{Goldman}, \binits{M.J.}},
\bauthor{\bsnm{Craft}, \binits{B.}},
\bauthor{\bsnm{Hastie}, \binits{M.}},
\bauthor{\bsnm{Repe{\v{c}}ka}, \binits{K.}},
\bauthor{\bsnm{McDade}, \binits{F.}},
\bauthor{\bsnm{Kamath}, \binits{A.}},
\bauthor{\bsnm{Banerjee}, \binits{A.}},
\bauthor{\bsnm{Luo}, \binits{Y.}},
\bauthor{\bsnm{Rogers}, \binits{D.}},
\bauthor{\bsnm{Brooks}, \binits{A.N.}}, \betal:
\batitle{Visualizing and interpreting cancer genomics data via the xena
  platform}.
\bjtitle{Nature biotechnology}
\bvolume{38}(\bissue{6}),
\bfpage{675}--\blpage{678}
(\byear{2020})
\end{barticle}
\endbibitem

\end{thebibliography}

\newcommand{\BMCxmlcomment}[1]{}

\BMCxmlcomment{

<refgrp>

<bibl id="B1">
  <title><p>Cancer molecular subtype classification by graph convolutional
  networks on multi-omics data</p></title>
  <aug>
    <au><snm>Li</snm><fnm>B</fnm></au>
    <au><snm>Wang</snm><fnm>T</fnm></au>
    <au><snm>Nabavi</snm><fnm>S</fnm></au>
  </aug>
  <source>Proceedings of the 12th ACM Conference on Bioinformatics,
  Computational Biology, and Health Informatics, BCB 2021</source>
  <publisher>Association for Computing Machinery</publisher>
  <pubdate>2021</pubdate>
  <volume>1</volume>
</bibl>

<bibl id="B2">
  <title><p>Integrated Multi-omics Analysis Using Variational Autoencoders:
  Application to Pan-cancer Classification</p></title>
  <aug>
    <au><snm>Zhang</snm><fnm>X</fnm></au>
    <au><snm>Zhang</snm><fnm>J</fnm></au>
    <au><snm>Sun</snm><fnm>K</fnm></au>
    <au><snm>Yang</snm><fnm>X</fnm></au>
    <au><snm>Dai</snm><fnm>C</fnm></au>
    <au><snm>Guo</snm><fnm>Y</fnm></au>
  </aug>
  <source>Proceedings - 2019 IEEE International Conference on Bioinformatics
  and Biomedicine, BIBM 2019</source>
  <pubdate>2019</pubdate>
  <fpage>765</fpage>
  <lpage>769</lpage>
</bibl>

<bibl id="B3">
  <title><p>Integrating Multi-Omic Data with Deep Subspace Fusion Clustering
  for Cancer Subtype Prediction</p></title>
  <aug>
    <au><snm>Yang</snm><fnm>B</fnm></au>
    <au><snm>Zhang</snm><fnm>Y</fnm></au>
    <au><snm>Pang</snm><fnm>S</fnm></au>
    <au><snm>Shang</snm><fnm>X</fnm></au>
    <au><snm>Zhao</snm><fnm>X</fnm></au>
    <au><snm>Han</snm><fnm>M</fnm></au>
  </aug>
  <source>IEEE/ACM Transactions on Computational Biology and
  Bioinformatics</source>
  <publisher>IEEE</publisher>
  <pubdate>2019</pubdate>
  <volume>XX</volume>
  <fpage>1</fpage>
  <lpage>1</lpage>
</bibl>

<bibl id="B4">
  <title><p>MOLI: Multi-omics late integration with deep neural networks for
  drug response prediction</p></title>
  <aug>
    <au><snm>Sharifi Noghabi</snm><fnm>H</fnm></au>
    <au><snm>Zolotareva</snm><fnm>O</fnm></au>
    <au><snm>Collins</snm><fnm>CC</fnm></au>
    <au><snm>Ester</snm><fnm>M</fnm></au>
  </aug>
  <source>Bioinformatics</source>
  <pubdate>2019</pubdate>
  <volume>35</volume>
  <fpage>i501</fpage>
  <lpage>i509</lpage>
</bibl>

<bibl id="B5">
  <title><p>MOGONET integrates multi-omics data using graph convolutional
  networks allowing patient classification and biomarker
  identification</p></title>
  <aug>
    <au><snm>Wang</snm><fnm>T</fnm></au>
    <au><snm>Shao</snm><fnm>W</fnm></au>
    <au><snm>Huang</snm><fnm>Z</fnm></au>
    <au><snm>Tang</snm><fnm>H</fnm></au>
    <au><snm>Zhang</snm><fnm>J</fnm></au>
    <au><snm>Ding</snm><fnm>Z</fnm></au>
    <au><snm>Huang</snm><fnm>K</fnm></au>
  </aug>
  <source>Nature Communications</source>
  <pubdate>2021</pubdate>
  <volume>12</volume>
  <fpage>3445</fpage>
  <url>http://www.nature.com/articles/s41467-021-23774-w</url>
</bibl>

<bibl id="B6">
  <title><p>Integrate multi-omics data with biological interaction networks
  using Multi-view Factorization AutoEncoder (MAE)</p></title>
  <aug>
    <au><snm>Ma</snm><fnm>T</fnm></au>
    <au><snm>Zhang</snm><fnm>A</fnm></au>
  </aug>
  <source>BMC Genomics</source>
  <publisher>BMC Genomics</publisher>
  <pubdate>2019</pubdate>
  <volume>20</volume>
  <fpage>1</fpage>
  <lpage>11</lpage>
  <url>http://dx.doi.org/10.1186/s12864-019-6285-x</url>
</bibl>

<bibl id="B7">
  <title><p>Multi-Omic Graph Transformers for Cancer Classification and
  Interpretation</p></title>
  <aug>
    <au><snm>Kaczmarek</snm><fnm>E</fnm></au>
    <au><snm>Jamzad</snm><fnm>A</fnm></au>
    <au><snm>Imtiaz</snm><fnm>T</fnm></au>
    <au><snm>Nanayakkara</snm><fnm>J</fnm></au>
    <au><snm>Renwick</snm><fnm>N</fnm></au>
    <au><snm>Mousavi</snm><fnm>P</fnm></au>
  </aug>
  <source>Pacific Symposium on Biocomputing. Pacific Symposium on
  Biocomputing</source>
  <pubdate>2022</pubdate>
  <volume>27</volume>
  <fpage>373</fpage>
  <lpage>384</lpage>
  <url>http://doi.org/10.1142/9789811250477_0034</url>
</bibl>

<bibl id="B8">
  <title><p>Multigrate : single-cell multi-omic data integration</p></title>
  <aug>
    <au><snm>Lotfollahi</snm><fnm>M</fnm></au>
    <au><snm>Litinetskaya</snm><fnm>A</fnm></au>
    <au><snm>Theis</snm><fnm>FJ</fnm></au>
  </aug>
  <pubdate>2022</pubdate>
  <fpage>1</fpage>
  <lpage>5</lpage>
</bibl>

<bibl id="B9">
  <title><p>Salmon: Survival analysis learning with multi-omics neural networks
  on breast cancer</p></title>
  <aug>
    <au><snm>Huang</snm><fnm>Z</fnm></au>
    <au><snm>Zhan</snm><fnm>X</fnm></au>
    <au><snm>Xiang</snm><fnm>S</fnm></au>
    <au><snm>Johnson</snm><fnm>TS</fnm></au>
    <au><snm>Helm</snm><fnm>B</fnm></au>
    <au><snm>Yu</snm><fnm>CY</fnm></au>
    <au><snm>Zhang</snm><fnm>J</fnm></au>
    <au><snm>Salama</snm><fnm>P</fnm></au>
    <au><snm>Rizkalla</snm><fnm>M</fnm></au>
    <au><snm>Han</snm><fnm>Z</fnm></au>
    <au><snm>Huang</snm><fnm>K</fnm></au>
  </aug>
  <source>Frontiers in Genetics</source>
  <pubdate>2019</pubdate>
  <volume>10</volume>
  <fpage>1</fpage>
  <lpage>13</lpage>
</bibl>

<bibl id="B10">
  <title><p>Semi-supervised classification of disease prognosis using CR images
  with clinical data structured graph</p></title>
  <aug>
    <au><snm>Bai</snm><fnm>J</fnm></au>
    <au><snm>Li</snm><fnm>B</fnm></au>
    <au><snm>Nabavi</snm><fnm>S</fnm></au>
  </aug>
  <source>Proceedings of the 13th ACM International Conference on
  Bioinformatics, Computational Biology and Health Informatics</source>
  <pubdate>2022</pubdate>
  <fpage>1</fpage>
  <lpage>-9</lpage>
</bibl>

<bibl id="B11">
  <title><p>Integrating multi-omics data through deep learning for accurate
  cancer prognosis prediction</p></title>
  <aug>
    <au><snm>Chai</snm><fnm>H</fnm></au>
    <au><snm>Zhou</snm><fnm>X</fnm></au>
    <au><snm>Zhang</snm><fnm>Z</fnm></au>
    <au><snm>Rao</snm><fnm>J</fnm></au>
    <au><snm>Zhao</snm><fnm>H</fnm></au>
    <au><snm>Yang</snm><fnm>Y</fnm></au>
  </aug>
  <source>Computers in biology and medicine</source>
  <publisher>Elsevier</publisher>
  <pubdate>2021</pubdate>
  <volume>134</volume>
  <fpage>104481</fpage>
</bibl>

<bibl id="B12">
  <title><p>Integrative multi-omics approaches in cancer research: From
  biological networks to clinical subtypes</p></title>
  <aug>
    <au><snm>Heo</snm><fnm>YJ</fnm></au>
    <au><snm>Hwa</snm><fnm>C</fnm></au>
    <au><snm>Lee</snm><fnm>GH</fnm></au>
    <au><snm>Park</snm><fnm>JM</fnm></au>
    <au><snm>An</snm><fnm>JY</fnm></au>
  </aug>
  <source>Molecules and Cells</source>
  <pubdate>2021</pubdate>
  <volume>44</volume>
  <fpage>433</fpage>
  <lpage>443</lpage>
</bibl>

<bibl id="B13">
  <title><p>Multiplatform analysis of 12 cancer types reveals molecular
  classification within and across tissues of origin</p></title>
  <aug>
    <au><snm>Hoadley</snm><fnm>KA</fnm></au>
    <au><snm>Yau</snm><fnm>C</fnm></au>
    <au><snm>Wolf</snm><fnm>DM</fnm></au>
    <au><snm>Cherniack</snm><fnm>AD</fnm></au>
    <au><snm>Tamborero</snm><fnm>D</fnm></au>
    <au><snm>Ng</snm><fnm>S</fnm></au>
    <au><snm>Leiserson</snm><fnm>MD</fnm></au>
    <au><snm>Niu</snm><fnm>B</fnm></au>
    <au><snm>McLellan</snm><fnm>MD</fnm></au>
    <au><snm>Uzunangelov</snm><fnm>V</fnm></au>
    <au><cnm>others</cnm></au>
  </aug>
  <source>Cell</source>
  <publisher>Elsevier</publisher>
  <pubdate>2014</pubdate>
  <volume>158</volume>
  <issue>4</issue>
  <fpage>929</fpage>
  <lpage>-944</lpage>
</bibl>

<bibl id="B14">
  <title><p>Delivering precision oncology to patients with cancer</p></title>
  <aug>
    <au><snm>Mateo</snm><fnm>J</fnm></au>
    <au><snm>Steuten</snm><fnm>L</fnm></au>
    <au><snm>Aftimos</snm><fnm>P</fnm></au>
    <au><snm>Andr{\'e}</snm><fnm>F</fnm></au>
    <au><snm>Davies</snm><fnm>M</fnm></au>
    <au><snm>Garralda</snm><fnm>E</fnm></au>
    <au><snm>Geissler</snm><fnm>J</fnm></au>
    <au><snm>Husereau</snm><fnm>D</fnm></au>
    <au><snm>Martinez Lopez</snm><fnm>I</fnm></au>
    <au><snm>Normanno</snm><fnm>N</fnm></au>
    <au><cnm>others</cnm></au>
  </aug>
  <source>Nature Medicine</source>
  <publisher>Nature Publishing Group</publisher>
  <pubdate>2022</pubdate>
  <volume>28</volume>
  <issue>4</issue>
  <fpage>658</fpage>
  <lpage>-665</lpage>
</bibl>

<bibl id="B15">
  <title><p>Cell-of-origin patterns dominate the molecular classification of
  10,000 tumors from 33 types of cancer</p></title>
  <aug>
    <au><snm>Hoadley</snm><fnm>KA</fnm></au>
    <au><snm>Yau</snm><fnm>C</fnm></au>
    <au><snm>Hinoue</snm><fnm>T</fnm></au>
    <au><snm>Wolf</snm><fnm>DM</fnm></au>
    <au><snm>Lazar</snm><fnm>AJ</fnm></au>
    <au><snm>Drill</snm><fnm>E</fnm></au>
    <au><snm>Shen</snm><fnm>R</fnm></au>
    <au><snm>Taylor</snm><fnm>AM</fnm></au>
    <au><snm>Cherniack</snm><fnm>AD</fnm></au>
    <au><snm>Thorsson</snm><fnm>V</fnm></au>
    <au><cnm>others</cnm></au>
  </aug>
  <source>Cell</source>
  <publisher>Elsevier</publisher>
  <pubdate>2018</pubdate>
  <volume>173</volume>
  <issue>2</issue>
  <fpage>291</fpage>
  <lpage>-304</lpage>
</bibl>

<bibl id="B16">
  <title><p>Convolutional neural networks on graphs with fast localized
  spectral filtering</p></title>
  <aug>
    <au><snm>Defferrard</snm><fnm>M</fnm></au>
    <au><snm>Bresson</snm><fnm>X</fnm></au>
    <au><snm>Vandergheynst</snm><fnm>P</fnm></au>
  </aug>
  <source>Advances in Neural Information Processing Systems</source>
  <pubdate>2016</pubdate>
  <fpage>3844</fpage>
  <lpage>3852</lpage>
</bibl>

<bibl id="B17">
  <title><p>A primer on deep learning in genomics</p></title>
  <aug>
    <au><snm>Zou</snm><fnm>J</fnm></au>
    <au><snm>Huss</snm><fnm>M</fnm></au>
    <au><snm>Abid</snm><fnm>A</fnm></au>
    <au><snm>Mohammadi</snm><fnm>P</fnm></au>
    <au><snm>Torkamani</snm><fnm>A</fnm></au>
    <au><snm>Telenti</snm><fnm>A</fnm></au>
  </aug>
  <source>Nature genetics</source>
  <publisher>Nature Publishing Group</publisher>
  <pubdate>2019</pubdate>
  <volume>51</volume>
  <issue>1</issue>
  <fpage>12</fpage>
  <lpage>-18</lpage>
</bibl>

<bibl id="B18">
  <title><p>Data-driven distributionally robust electric vehicle balancing for
  mobility-on-demand systems under demand and supply uncertainties</p></title>
  <aug>
    <au><snm>He</snm><fnm>S</fnm></au>
    <au><snm>Pepin</snm><fnm>L</fnm></au>
    <au><snm>Wang</snm><fnm>G</fnm></au>
    <au><snm>Zhang</snm><fnm>D</fnm></au>
    <au><snm>Miao</snm><fnm>F</fnm></au>
  </aug>
  <source>2020 IEEE/RSJ International Conference on Intelligent Robots and
  Systems (IROS)</source>
  <fpage>2165</fpage>
  <lpage>-2172</lpage>
</bibl>

<bibl id="B19">
  <title><p>Single-cell RNA sequencing data clustering using graph
  convolutional networks</p></title>
  <aug>
    <au><snm>Wang</snm><fnm>T</fnm></au>
    <au><snm>Li</snm><fnm>B</fnm></au>
    <au><snm>Nabavi</snm><fnm>S</fnm></au>
  </aug>
  <source>2021 IEEE International Conference on Bioinformatics and Biomedicine
  (BIBM)</source>
  <pubdate>2021</pubdate>
  <fpage>2163</fpage>
  <lpage>-2170</lpage>
</bibl>

<bibl id="B20">
  <title><p>Constraint Learning to Define Trust Regions in Predictive-Model
  Embedded Optimization</p></title>
  <aug>
    <au><snm>Shi</snm><fnm>C</fnm></au>
    <au><snm>Emadikhiav</snm><fnm>M</fnm></au>
    <au><snm>Lozano</snm><fnm>L</fnm></au>
    <au><snm>Bergman</snm><fnm>D</fnm></au>
  </aug>
  <source>arXiv preprint arXiv:2201.04429</source>
  <pubdate>2022</pubdate>
</bibl>

<bibl id="B21">
  <title><p>Robust electric vehicle balancing of autonomous mobility-on-demand
  system: A multi-agent reinforcement learning approach</p></title>
  <aug>
    <au><snm>He</snm><fnm>S</fnm></au>
    <au><snm>Han</snm><fnm>S</fnm></au>
    <au><snm>Miao</snm><fnm>F</fnm></au>
  </aug>
  <source>2023 IEEE/RSJ International Conference on Intelligent Robots and
  Systems (IROS)</source>
  <pubdate>2023</pubdate>
  <fpage>5471</fpage>
  <lpage>-5478</lpage>
</bibl>

<bibl id="B22">
  <title><p>A two-stage exact algorithm for optimization of neural network
  ensemble</p></title>
  <aug>
    <au><snm>Wang</snm><fnm>K</fnm></au>
    <au><snm>Lozano</snm><fnm>L</fnm></au>
    <au><snm>Bergman</snm><fnm>D</fnm></au>
    <au><snm>Cardonha</snm><fnm>C</fnm></au>
  </aug>
  <source>Integration of Constraint Programming, Artificial Intelligence, and
  Operations Research: 18th International Conference, CPAIOR 2021, Vienna,
  Austria, July 5--8, 2021, Proceedings 18</source>
  <pubdate>2021</pubdate>
  <fpage>106</fpage>
  <lpage>-114</lpage>
</bibl>

<bibl id="B23">
  <title><p>Integrated multi-omics analyses in oncology: a review of machine
  learning methods and tools</p></title>
  <aug>
    <au><snm>Nicora</snm><fnm>G</fnm></au>
    <au><snm>Vitali</snm><fnm>F</fnm></au>
    <au><snm>Dagliati</snm><fnm>A</fnm></au>
    <au><snm>Geifman</snm><fnm>N</fnm></au>
    <au><snm>Bellazzi</snm><fnm>R</fnm></au>
  </aug>
  <source>Frontiers in oncology</source>
  <publisher>Frontiers Media SA</publisher>
  <pubdate>2020</pubdate>
  <volume>10</volume>
  <fpage>1030</fpage>
</bibl>

<bibl id="B24">
  <title><p>A comprehensive survey on graph neural networks</p></title>
  <aug>
    <au><snm>Wu</snm><fnm>Z</fnm></au>
    <au><snm>Pan</snm><fnm>S</fnm></au>
    <au><snm>Chen</snm><fnm>F</fnm></au>
    <au><snm>Long</snm><fnm>G</fnm></au>
    <au><snm>Zhang</snm><fnm>C</fnm></au>
    <au><snm>Philip</snm><fnm>SY</fnm></au>
  </aug>
  <source>IEEE transactions on neural networks and learning systems</source>
  <publisher>IEEE</publisher>
  <pubdate>2020</pubdate>
</bibl>

<bibl id="B25">
  <title><p>Graph attention networks</p></title>
  <aug>
    <au><snm>Velicković</snm><fnm>P</fnm></au>
    <au><snm>Cucurull</snm><fnm>G</fnm></au>
    <au><snm>Casanova</snm><fnm>A</fnm></au>
    <au><snm>Romero</snm><fnm>A</fnm></au>
    <au><snm>Liò</snm><fnm>P</fnm></au>
    <au><snm>Bengio</snm><fnm>Y</fnm></au>
  </aug>
  <source>arXiv</source>
  <pubdate>2017</pubdate>
  <fpage>1</fpage>
  <lpage>12</lpage>
</bibl>

<bibl id="B26">
  <title><p>Classification of Cancer Types Using Graph Convolutional Neural
  Networks</p></title>
  <aug>
    <au><snm>Ramirez</snm><fnm>R</fnm></au>
    <au><snm>Chiu</snm><fnm>YC</fnm></au>
    <au><snm>Hererra</snm><fnm>A</fnm></au>
    <au><snm>Mostavi</snm><fnm>M</fnm></au>
    <au><snm>Ramirez</snm><fnm>J</fnm></au>
    <au><snm>Chen</snm><fnm>Y</fnm></au>
    <au><snm>Huang</snm><fnm>Y</fnm></au>
    <au><snm>Jin</snm><fnm>YF</fnm></au>
  </aug>
  <source>Frontiers in physics</source>
  <publisher>NIH Public Access</publisher>
  <pubdate>2020</pubdate>
  <volume>8</volume>
</bibl>

<bibl id="B27">
  <title><p>Single-cell classification using graph convolutional
  networks</p></title>
  <aug>
    <au><snm>Wang</snm><fnm>T</fnm></au>
    <au><snm>Bai</snm><fnm>J</fnm></au>
    <au><snm>Nabavi</snm><fnm>S</fnm></au>
  </aug>
  <source>BMC bioinformatics</source>
  <publisher>BioMed Central</publisher>
  <pubdate>2021</pubdate>
  <volume>22</volume>
  <issue>1</issue>
  <fpage>1</fpage>
  <lpage>-23</lpage>
</bibl>

<bibl id="B28">
  <title><p>GrAMME: Semisupervised Learning Using Multilayered Graph Attention
  Models</p></title>
  <aug>
    <au><snm>Shanthamallu</snm><fnm>US</fnm></au>
    <au><snm>Thiagarajan</snm><fnm>JJ</fnm></au>
    <au><snm>Song</snm><fnm>H</fnm></au>
    <au><snm>Spanias</snm><fnm>A</fnm></au>
  </aug>
  <source>IEEE Transactions on Neural Networks and Learning Systems</source>
  <pubdate>2020</pubdate>
  <volume>31</volume>
  <fpage>3977</fpage>
  <lpage>3988</lpage>
</bibl>

<bibl id="B29">
  <title><p>Breast cancer subtypes based on ER/PR and Her2 expression:
  comparison of clinicopathologic features and survival</p></title>
  <aug>
    <au><snm>Onitilo</snm><fnm>AA</fnm></au>
    <au><snm>Engel</snm><fnm>JM</fnm></au>
    <au><snm>Greenlee</snm><fnm>RT</fnm></au>
    <au><snm>Mukesh</snm><fnm>BN</fnm></au>
  </aug>
  <source>Clinical medicine \& research</source>
  <publisher>Marshfield Clinic</publisher>
  <pubdate>2009</pubdate>
  <volume>7</volume>
  <issue>1-2</issue>
  <fpage>4</fpage>
  <lpage>-13</lpage>
</bibl>

<bibl id="B30">
  <title><p>{{T}he {B}io{G}{R}{I}{D} database: {A} comprehensive biomedical
  resource of curated protein, genetic, and chemical interactions}</p></title>
  <aug>
    <au><snm>Oughtred</snm><fnm>R.</fnm></au>
    <au><snm>Rust</snm><fnm>J.</fnm></au>
    <au><snm>Chang</snm><fnm>C.</fnm></au>
    <au><snm>Breitkreutz</snm><fnm>B. J.</fnm></au>
    <au><snm>Stark</snm><fnm>C.</fnm></au>
    <au><snm>Willems</snm><fnm>A.</fnm></au>
    <au><snm>Boucher</snm><fnm>L.</fnm></au>
    <au><snm>Leung</snm><fnm>G.</fnm></au>
    <au><snm>Kolas</snm><fnm>N.</fnm></au>
    <au><snm>Zhang</snm><fnm>F.</fnm></au>
    <au><snm>Dolma</snm><fnm>S.</fnm></au>
    <au><snm>Coulombe Huntington</snm><fnm>J.</fnm></au>
    <au><snm>Chatr Aryamontri</snm><fnm>A.</fnm></au>
    <au><snm>Dolinski</snm><fnm>K.</fnm></au>
    <au><snm>Tyers</snm><fnm>M.</fnm></au>
  </aug>
  <source>Protein Sci</source>
  <pubdate>2021</pubdate>
  <volume>30</volume>
  <issue>1</issue>
  <fpage>187</fpage>
  <lpage>-200</lpage>
</bibl>

<bibl id="B31">
  <title><p>miRDB: an online database for prediction of functional microRNA
  targets</p></title>
  <aug>
    <au><snm>Chen</snm><fnm>Y</fnm></au>
    <au><snm>Wang</snm><fnm>X</fnm></au>
  </aug>
  <source>Nucleic acids research</source>
  <publisher>Oxford University Press</publisher>
  <pubdate>2020</pubdate>
  <volume>48</volume>
  <issue>D1</issue>
  <fpage>D127</fpage>
  <lpage>-D131</lpage>
</bibl>

<bibl id="B32">
  <title><p>Heterogeneous Multi-Layered Network Model for Omics Data
  Integration and Analysis</p></title>
  <aug>
    <au><snm>Lee</snm><fnm>B</fnm></au>
    <au><snm>Zhang</snm><fnm>S</fnm></au>
    <au><snm>Poleksic</snm><fnm>A</fnm></au>
    <au><snm>Xie</snm><fnm>L</fnm></au>
  </aug>
  <source>Frontiers in Genetics</source>
  <pubdate>2020</pubdate>
  <volume>10</volume>
  <fpage>1</fpage>
  <lpage>11</lpage>
</bibl>

<bibl id="B33">
  <title><p>Comprehensive molecular portraits of human breast
  tumours</p></title>
  <aug>
    <au><snm>13</snm><fnm>BWHHMSCLPPJKR</fnm></au>
    <au><snm>Medicine Creighton Chad J. 22 23 Donehower Lawrence A. 22 23 24
  25</snm><fnm>G</fnm></au>
    <au><snm>Systems Biology Reynolds Sheila 31 Kreisberg Richard B. 31 Bernard
  Brady 31 Bressler Ryan 31 Erkkila Timo 32 Lin Jake 31 Thorsson Vesteinn 31
  Zhang Wei 33 Shmulevich Ilya 31</snm><fnm>I</fnm></au>
    <au><cnm>others</cnm></au>
  </aug>
  <source>Nature</source>
  <publisher>Nature Publishing Group UK London</publisher>
  <pubdate>2012</pubdate>
  <volume>490</volume>
  <issue>7418</issue>
  <fpage>61</fpage>
  <lpage>-70</lpage>
</bibl>

<bibl id="B34">
  <title><p>Visualizing and interpreting cancer genomics data via the Xena
  platform</p></title>
  <aug>
    <au><snm>Goldman</snm><fnm>MJ</fnm></au>
    <au><snm>Craft</snm><fnm>B</fnm></au>
    <au><snm>Hastie</snm><fnm>M</fnm></au>
    <au><snm>Repe{\v{c}}ka</snm><fnm>K</fnm></au>
    <au><snm>McDade</snm><fnm>F</fnm></au>
    <au><snm>Kamath</snm><fnm>A</fnm></au>
    <au><snm>Banerjee</snm><fnm>A</fnm></au>
    <au><snm>Luo</snm><fnm>Y</fnm></au>
    <au><snm>Rogers</snm><fnm>D</fnm></au>
    <au><snm>Brooks</snm><fnm>AN</fnm></au>
    <au><cnm>others</cnm></au>
  </aug>
  <source>Nature biotechnology</source>
  <publisher>Nature Publishing Group</publisher>
  <pubdate>2020</pubdate>
  <volume>38</volume>
  <issue>6</issue>
  <fpage>675</fpage>
  <lpage>-678</lpage>
</bibl>

</refgrp>
} 









\section*{Additional Files}
  \subsection*{Additional file 1: Detailed Results and Model Settings}

\end{backmatter}
\end{document}